\documentclass{article}
\usepackage{amssymb,amsmath,epsfig,epstopdf,color,graphicx,subfigure}
\usepackage{epstopdf,array}
\newcolumntype{L}{>{\centering\arraybackslash}m{3cm}}

\usepackage{arxiv}

\usepackage[utf8]{inputenc} 
\usepackage[T1]{fontenc}    
\usepackage{hyperref}       
\usepackage{url}            
\usepackage{booktabs}       
\usepackage{amsfonts}       
\usepackage{nicefrac}       
\usepackage{microtype}      

\usepackage{amsmath,amssymb,amsfonts}
\usepackage{algorithmic}
\usepackage{graphicx}
\usepackage{textcomp}
\usepackage{comment,color}
\usepackage{enumitem}
\usepackage{epsfig}
\usepackage{caption}
\usepackage{verbatim,bm}
\usepackage{epstopdf,array,mathtools,url,slashbox,subfigure}

\newcolumntype{L}{>{\centering\arraybackslash}m{3cm}}
\DeclarePairedDelimiter\norm{\lVert}{\rVert}%
\makeatletter
\let\oldabs\abs
\def\abs{\@ifstar{\oldabs}{\oldabs*}}
\let\oldnorm\norm
\def\norm{\@ifstar{\oldnorm}{\oldnorm*}}
\makeatother
\title{A new practical and effective source-independent full-waveform inversion with a velocity-distribution supported deep image prior: Applications to two real datasets}

\author{
  Chao Song\\
 Department of Geophysics, College of Geo-\\
 Exploration Science and Technology, \\
 Jilin University
    \And
   Tariq Alkhalifah\\
Physical Sciences and Engineering Division, \\
 King Abdullah University of Science and Technology (KAUST),
    \And 
  Umair bin Waheed\\
  Department of Geosciences\\
  King Fahd University of Petroleum and Minerals
    \And  
  Silin Wang\\
 Department of Geophysics, College of Geo-\\
 Exploration Science and Technology, \\
 Jilin University
    \And  
  Cai Liu\\
 Department of Geophysics, College of Geo-\\
 Exploration Science and Technology, \\
 Jilin University
    \And  
}

\begin{document}
\maketitle

\begin{abstract}

Full-waveform inversion (FWI) is an advanced technique for reconstructing high-resolution subsurface physical parameters by progressively minimizing the discrepancy between observed and predicted seismic data. However, conventional FWI encounters challenges in real data applications, primarily due to its conventional objective of direct measurements of the data misfit. Accurate estimation of the source wavelet is essential for effective data fitting, alongside the need for low-frequency data and a reasonable initial model to prevent cycle skipping. Additionally, wave equation solvers often struggle to accurately simulate the amplitude of observed data in real applications. To address these challenges, we introduce a correlation-based source-independent objective function for FWI that aims to mitigate source uncertainty and amplitude dependency, which effectively enhances its practicality for real data applications. We develop a deep-learning framework constrained by this new objective function with a velocity-distribution supported deep image prior, which reparameterizes velocity inversion into trainable parameters within an autoencoder, thereby reducing the nonlinearity in the conventional FWI's objective function. We demonstrate the superiority of our proposed method using synthetic data from benchmark velocity models and, more importantly, two real datasets. These examples highlight its effectiveness and practicality even under challenging conditions, such as missing low frequencies, a crude initial velocity model, and an incorrect source wavelet.

\end{abstract}


\section{Introduction}

Seismic imaging plays a crucial role in exploring and characterizing subsurface structures, providing valuable insights across fields such as exploring natural resources and imaging the Earth's interior. Among various seismic imaging techniques, full-waveform inversion (FWI) stands out for its ability to reconstruct high-resolution subsurface models by minimizing the misfit between observed and predicted seismic data utilizing all types of seismic waves \cite{tarantola1984inversion,brenders2007full,virieux2009overview,wang2009reflection,sirgue2010thematic,alkhalifah2016full}. While the concept of FWI is relatively straightforward, it faces significant limitations that can hinder its effectiveness in practical applications.

One major challenge is the inherent high nonlinearity of the objective function. When the difference in arrival times between observed and predicted seismic data exceeds half a cycle, FWI can get stuck in a local minimum, a phenomenon known as cycle skipping \cite{virieux2009overview}. Though low-frequency data and a good initial model can mitigate this adverse effect, these factors are often unavailable in realistic FWI applications. To address this, several techniques have been developed to recover low-frequency data and construct more reliable background velocity models. These methods include seismic traveltime-based inversion \cite{luo1991wave, leung2006adjoint}, migration-based velocity analysis \cite{sava2004wave, symes2008migration}, envelope-based inversion \cite{chi2014full, wu2014seismic}, and reflection-based waveform inversion \cite{xu2012full, wu2015simultaneous, yao2020review}, etc. These approaches prioritize obtaining a good initial velocity model, which requires a sequential FWI implementation to achieve the final high-resolution velocity model.

Additionally, researchers have developed direct solutions to the cycle-skipping issue by creating new objective functions to measure discrepancies between observed and predicted data. Approaches such as adaptive waveform inversion (AWI) \cite{warner2016adaptive, guasch2019adaptive}, wavefield reconstruction inversion \cite{van2013mitigating, van2015penalty, song2020efficient}, and optimal transport \cite{engquist2016optimal, yang2018application} offer alternative formulations that enhance the robustness of the inversion process. These methods aim to minimize the misfit more effectively, reducing the sensitivity to initial conditions and improving overall stability. While these advancements significantly enhance FWI, yet challenges remain.

FWI heavily relies on accurate source wavelet estimation, as uncertainty in the source can significantly affect the quality of the inversion results. To address this issue, source-independent FWI (SIFWI) was proposed to measure the misfit of convolved data \cite{choi2011source}. The convolved data are given by the observed data convolved with a reference trace from the predicted data, and vice versa. SIFWI effectively eliminates the source wavelet effect. This feature allows for the use of this objective function in passive seismic inversion \cite{wang2018microseismic}. However, amplitude discrepancy between simulated and measured data may also pose a challenge as conventional wave equation solvers may struggle to accurately simulate the amplitude of seismic waves during propagation. In practice, the amplitude in the observed data can be distorted by data processing and receiver coupling, leading to discrepancies in the inversion process. \cite{choi2012application} developed a global correlation norm (GCN) based objective function to mitigate amplitude dependency, enhancing the practicality of FWI in real applications. From previous works, it is evident that various objective function formulations for FWI have been developed to tackle specific challenges, most relying on the adjoint-state method to derive the gradient of the target parameters \cite{plessix2006review}. However, merging these objective functions to address the respective challenges simultaneously is not trivial.

Deep learning (DL) has emerged as a revolutionary technique in scientific and engineering fields and has become a powerful tool for enhancing FWI \cite{karniadakis2021physics,wu2023sensing,schuster2024review}. Automatic differentiation in DL-based framework currently is gaining wide attention in FWI by enabling efficient computation of gradients, which facilitates the optimization of complex objective functions \cite{baydin2018automatic, richardson2018seismic, zhu2021general, song2023weighted}. This capability is particularly beneficial in addressing the nonlinear nature of FWI, allowing for more effective updates to model parameters during the inversion process. DL-based approaches can directly build the mapping between observed data and the subsurface model with neural networks (NNs), targeting the optimization parameters from the subsurface model to the network's trainable parameters. These approaches for FWI can be mainly categorized into data-driven and physics-constrained frameworks. Data-driven DL-based FWI methods rely on large datasets to learn mapping functions directly from the data. However, their accuracy and generalization ability are limited by the diversity of the training data \cite{yang2019deep, wu2019inversionnet, deng2022openfwi,kazei2021mapping}, making them difficult to apply to real datasets. In contrast, physics-constrained approaches incorporate physical principles to guide the learning process and do not require extensive training data. Though these methods may not be as efficient as data-driven approaches, they ensure that the inverted subsurface models remain consistent with geophysical laws. For instance, \cite{zhu2022integrating} used a randomized vector as input to a generative NN to represent velocity perturbations. The spatial coordinates were also employed as input to fully connected NNs to predict the velocity model, which is a common setup in physics-informed NNs (PINNs) \cite{waheed2021pinntomo,rasht2022physics,song2021wavefield,sun2023implicit, zhang2023multilayer}. Researchers have also utilized generative adversarial networks and autoencoders to map target velocity models from observed data \cite{yang2023fwigan, mardan2024physics}. Additionally, NNs have been employed as intelligent regularization tools to enhance FWI's performance \cite{zhang2020high, sun2023full, yao2023regularization}. However, physics-constrained DL-based FWI methods are often designed for specific datasets, limiting their generalization ability. 

In this paper, we develop a novel correlation-based SIFWI to fundamentally mitigate source uncertainty and amplitude dependency issues in FWI, which makes it more practical for real data applications. We incorporate a new objective function into a physics-constrained DL-based FWI framework given by an autoencoder. We propose using a group of velocity models from the OpenFWI dataset as the input into the autoencoder to form a velocity-distribution supported deep image prior to map the target output velocity model. OpenFWI velocity models can provide the autoencoder with valuable underlying geological features to benefit the prediction of the velocity model. We test our approach on both synthetic and real data, and analyze the results. 

The rest of the paper is organized as follows: we first start by introducing the source-independent correlation-based FWI framework. Then, we introduce the physical constraints to the implementation. This is followed by testing our developed approach on the Marmousi and the Overthrust models.  We then share applications on marine data from North West Australia and ocean-bottom cable (OBC) data from the North Sea. We finally summarize our observations and potential limitations in the discussion and follow that up with our conclusions.

\section{Theory}

\subsection{Correlation-based SIFWI}
We simulate the propagation of seismic waves by solving a constant-density acoustic wave equation, given by:
\begin{equation}
\frac{1}{v(\textbf{x})^{2}}\frac{\partial^{2} u_{s}(\textbf{x},t,v)}{\partial t^{2}}-\nabla^{2}u_{s}(\textbf{x},t,v)=w(t)\delta(\textbf{x}-\textbf{x}_{s}),
\label{eqn:eq1}
\end{equation}
where $u_{s}$ represents the seismic wavefield, $v$ represents the velocity model, and $w(t)$ represents the source wavelet. $\delta(\textbf{x}-\textbf{x}_{s})$ is the Dirac delta function, where $\textbf{x}$ and $\textbf{x}_{s}$ denote the model space and source coordinates, respectively. The indices $s$ and $t$ are used to represent the source and time, respectively. 

Conventional FWI minimizes the $l_{2}$ norm misfit between the observed and predicted data, expressed as:
\begin{equation}
J_{FWI}(v)=\underset{v}{\textbf{min}}\frac{1}{2}\sum_{s}\sum_{r}\int_{t}\left \| u_{s}(\textbf{x}_{r},t,v)-d_{s}(\textbf{x}_{r},t) \right \|_{2}^{2}dt,
\label{eqn:eq2}
\end{equation}
where $d_{s}(\textbf{x}_{r},t)$ represents the observed data for the source $s$, and $u_{s}(\textbf{x}_{r},t,v)$ represents the predicted data obtained by sampling the simulated seismic wavefield $u_{s}$ at receivers' coordinates ($\textbf{x}_{r}$). Here, $r$ denotes the index of the receivers. The notation $\left\|  \right\|_{2}$ denotes the $l_{2}$ norm. This objective function of FWI relies on a direct amplitude and phase comparison between observed and predicted data. However, solving the acoustic wave equation may not accurately simulate the amplitude variation of seismic waves during propagation. Additionally, the amplitude information of the observed data may be unreliable due to factors such as the elastic nature of the Earth, receiver coupling and data processing. Thus, it is crucial to eliminate amplitude dependency from FWI. To address this, a correlation-based FWI objective function based on the global correlation norm (GCN) was proposed \cite{choi2012application}, which is expressed as:

\begin{equation}
J_{GCN}(v)=\underset{v}{\textbf{min}}\sum_{s}\sum_{r}\int_{t}\left \{ - \widehat{u_{s}}(\textbf{x}_{r},t,v)\cdot \widehat{d_{s}}(\textbf{x}_{r},t)\right \} dt,
\label{eqn:eq3}
\end{equation}

where $\widehat{d}_{s}(\textbf{x}_{r},t)=\frac{d_{s}(\textbf{x}_{r},t)}{\left \| d_{s}(\textbf{x}_{r},t) \right \|}_{2}$ and $\widehat{u_{s}}(\textbf{x}_{r},t,v)=\frac{u_{s}(\textbf{x}_{r},t,v)}{\left \| u_{s}(\textbf{x}_{r},t,v) \right \|}_{2}$ denote the normalized observed and predicted data, respectively. The operation $\cdot$ represents the correlation operation. Though this GCN objective function effectively eliminates the dependency of FWI on data amplitude, it still requires data fitting for phase information. An accurate estimation of the source wavelet is a prerequisite for generating predicted data that fits the observed data in terms of phase. However, estimating the source wavelet from real data is challenging, as it is often contaminated by noise and it depends highly on the complexity of the near surface. To mitigate inversion errors caused by inaccurate source wavelet estimation, \cite{choi2011source} proposed a source-independent FWI based on the discrepancy of convolved wavefields. By convolving a reference trace from the predicted data with the observed data, and vice versa, the effect of the source wavelet phase difference is primely mitigated.

We embed the concept of source-independent FWI using convolved wavefields in the GCN objective function to create a new form of FWI, namely correlation-based SIFWI. This new objective function of FWI is not affected by inaccuracy of data amplitude and source wavelet, which is expressed as:
\begin{equation}
J_{new}(v)=\underset{v}{\textbf{min}}\sum_{s}\sum_{r}\int_{t}\left \{-\frac{ u_{s}(\textbf{x}_{r},t,v)\ast  d_{s}(x_{k},t)}{\left\|  u_{s}(\textbf{x}_{r},t,v)\ast d_{s}(x_{k},t) \right\|_{2}}\cdot \frac{ d_{s}(\textbf{x}_{r},t)\ast u_{s}(x_{k},t,v)}{\left\|  d_{s}(\textbf{x}_{r},t)\ast u_{s}(x_{k},t,v)\right\|_{2}}\right \} dt,
\label{eqn:eq4}
\end{equation}
where $x_{k}$ denotes the coordinate of the reference trace; The operation $\ast$ represents the convolution operation. The adjoint source for this objective function is complicated. However, through the utilization of automatic differentiation, as we will see next, the process of computing the gradient is both straightforward and automatically optimized.

\subsection{Physics-constrained DL-SIFWI}

Though the proposed correlation-based SIFWI effectively eliminates the data amplitude dependency and source uncertainty, it does not address the inherent nonlinearity in FWI, which remains a significant challenge. In conventional FWI implementations, the model space is given by a gridded representation of the velocity. With this representation of the model we do not allow for any extensions that may help avoid the inherent nonlinearity of FWI \cite{wu2017efficient,symes2020wavefield}. According to the theorem of universal approximation, NNs can approximate any target function \cite{hornik1991approximation,leshno1993multilayer}. In DL-based FWI, we can utilize an NN to represent the velocity model, reparameterizing the model space in FWI from velocity to trainable parameters within the network. This allows for an extension in the model space that we help us avoid cycle skipping. This feature has been demonstrated by prior studies \cite{yang2019deep,deng2022openfwi,dhara2022physics,yang2023fwigan,mardan2024physics}. Since the update of such an NN representation of the velocity is performed through its output velocity, there is no reason to have the observed data as input, as done in these studies, and thus, we refrain from doing so here. 

\begin{figure}
\begin{center}
\includegraphics[width=1.0\textwidth]{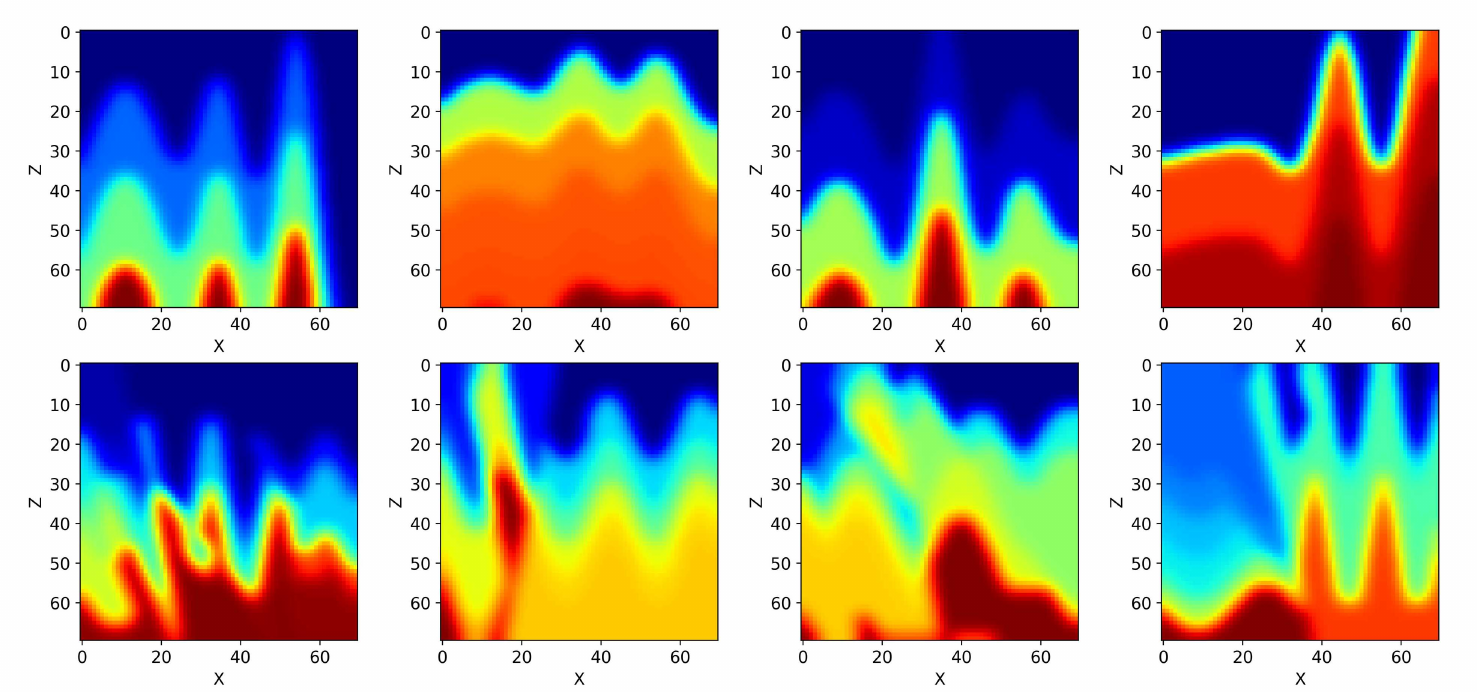} 
\caption{Selected smoothed velocity models from the CurveVel-B and CurveFault-B families in OpenFWI datasets.}
\label{fig:model_openfwi}
\end{center}   
\end{figure}

Distinguished from previous work, we alternatively propose to use a set of random velocity models as input. These models, for our tests, are sourced from the CurveVel-B and CurveFault-B families in the OpenFWI datasets \cite{deng2022openfwi}, with selected examples shown in Figure \ref{fig:model_openfwi.}. The reason we use them as input in DL-SIFWI is their small-size and availability, making them ideal for use in DL-SIFWI. 

\begin{figure}
\begin{center}
\includegraphics[width=1.0\textwidth]{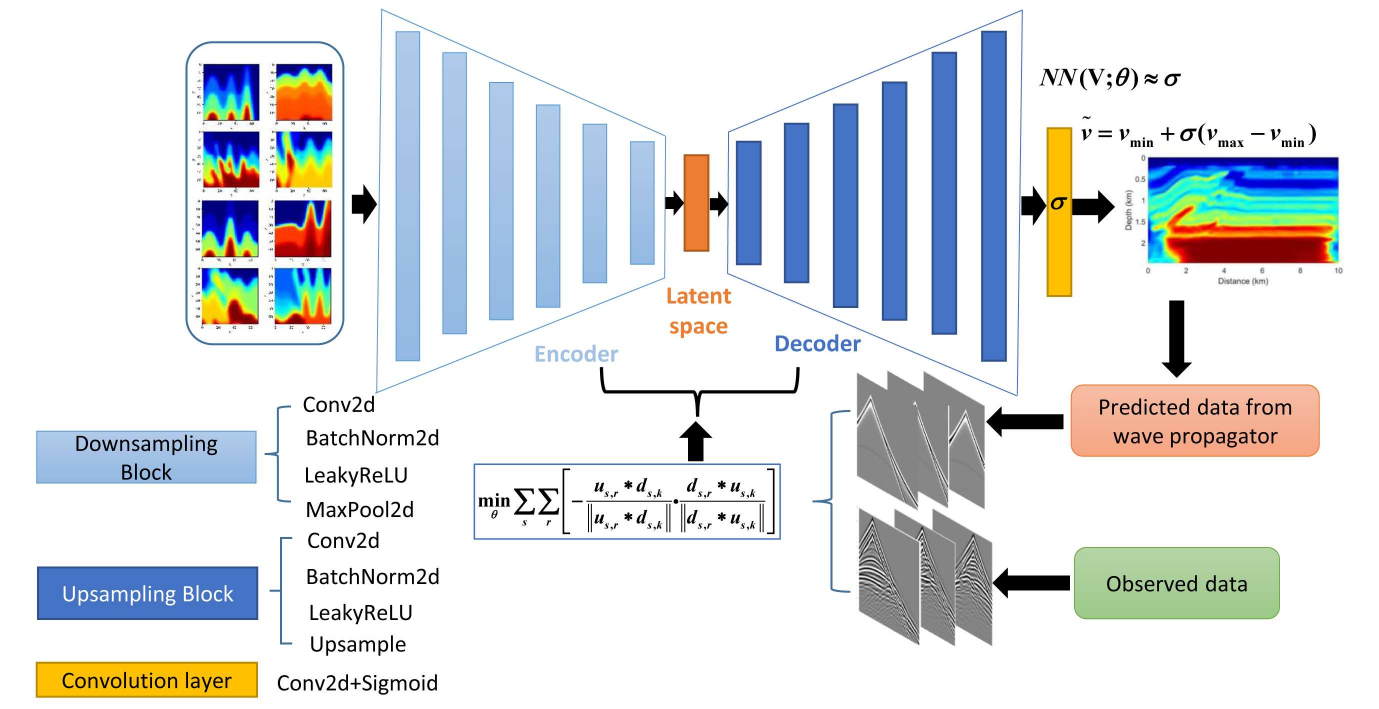} 
\caption{Schematic workflow of the proposed DL-SIFWI method. The input consists of 40 pieces of OpenFWI velocity models. The output is the inverted velocity model reparameterized by an autoencoder network. The loss function is the correlation of the convolved seismic data.}
\label{fig:Autoencoder}
\end{center}   
\end{figure}

We employ an autoencoder to process these velocity groups and predict the target velocity under the physics constraint of the proposed correlation-based SIFWI, mathematically expressed as $NN(\textbf{V};\theta)\approx \sigma$. Here, the variable $\textbf{V}$ represents the input velocity group; the parameter $\theta$ represents the trainable parameters in the autoencoder; the variable $\sigma$ represents the predicted scaled velocity. The autoencoder comprises six encoder blocks followed by an equal number of decoder blocks. Each encoder block includes a series of operations such as convolution, batch normalization, LeakyReLU activation, and max pooling, while upsampling blocks replace max pooling with the upsampling operation. Before outputting the scaled velocity $\sigma$, a convolution layer is applied, followed by a sigmoid activation to ensure that $\sigma$ is constrained between 0 and 1. The final predicted velocity is computed using the prior information of the minimum $v_{min}$ and maximum $v_{max}$ velocities, resulting in $\widetilde{v}=v_{min} + \sigma(v_{max}-v_{min})$. This predicted velocity is then utilized in a wave equation propagator, specifically from Deepwave \cite{richardson_alan_2022}, to generate the predicted data. Ultimately, the proposed correlation-based SIFWI loss function is employed to update the trainable parameters $\theta$ in the autoencoder. This DL-based SIFWI is abbreviated as DL-SIFWI and its workflow is illustrated in Figure \ref{fig:Autoencoder}.

\begin{equation}
J_{DL}(\theta)=\underset{\theta}{\textbf{min}}\sum_{s}\sum_{r}\int_{t}\left \{-\frac{ u_{s}(\textbf{x}_{r},t,\widetilde{v})\ast  d_{s}(x_{k},t)}{\left\|   u_{s}(\textbf{x}_{r},t,\widetilde{v})\ast d_{s}(x_{k},t) \right\|_{2}}\cdot \frac{ d_{s}(\textbf{x}_{r},t)\ast u_{s}(x_{k},t,\widetilde{v})}{\left\|  d_{s}(\textbf{x}_{r},t)\ast u_{s}(x_{k},t,\widetilde{v})\right\|_{2}}\right \} dt+\alpha\left\|  \widetilde{v}-v_{0}\right\|_{2}^{2}.
\label{eqn:eq5}
\end{equation}
We add a regularization term to equation \ref{eqn:eq5} to stabilize the training process if the initial velocity $v_{0}$ model is available, in which $\alpha$ is the weight factor that controls the regularization term. $\alpha$ varies for different epochs. As the training of DL-SIFWI progresses, we decrease the value of $\alpha$ to enhance the data-fitting objective.

The proposed DL-SIFWI method makes significant contributions in two key areas: first, it mitigates FWI's dependence on data amplitude and the source wavelet by introducing a novel objective function; second, it tackles the nonlinearity of FWI and overcomes the cycle-skipping issue by integrating this objective function into a physics-constrained DL framework.

\section{Results}

In this section, we apply the proposed DL-SIFWI method to synthetic data generated from the benchmark Marmousi and Overthrust velocity models, as well as to two real datasets. We demonstrate the method's superiority, particularly under conditions of extreme limitation in prior information, such as the absence of an initial velocity model, low-frequency data, and source wavelet information. 

\subsection{Marmousi model}

\begin{figure}
\begin{center}
\includegraphics[width=1.0\textwidth]{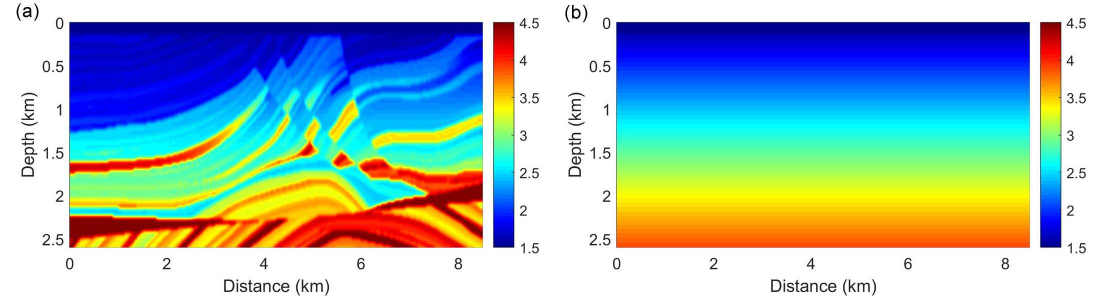} 
\caption{The (a) true Marmousi velocity model and the (b) initial velocity model.}
\label{fig:mar_true_init}
\end{center}   
\end{figure}

We begin our test with the Marmousi model using a grid size of  $131 \times 426$ and setting the sampling interval to 20 m. We show the true and initial velocity models in Figures \ref{fig:mar_true_init}a and \ref{fig:mar_true_init}b, respectively. The initial velocity model is a constant gradient model that lacks all detailed structural information.

\begin{figure}
\begin{center}
\includegraphics[width=1.0\textwidth]{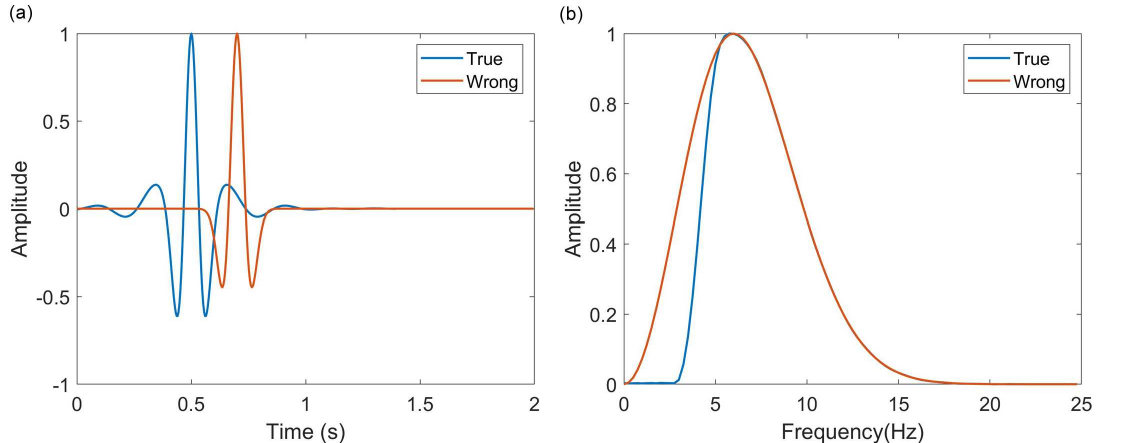} 
\caption{(a) The true and wrong wavelets and (b) their corresponding spectra. (Blue line: true wavelet; red line: wrong wavelet.)}
\label{fig:wavelet_spectra}
\end{center}   
\end{figure}

To generate the observed data, we consider a Ricker wavelet (dominant frequency: 6 Hz) without low-frequency components below 3 Hz as the true source wavelet (blue line in Figure \ref{fig:wavelet_spectra}a). We place 40 sources on the surface distributed from 0.1 km to 8.1 km at 200-m interval, with 426 receivers recording the observed data at a depth of 20 m. Since the true wavelet is unknown, we utilize a shifted 6-Hz Ricker wavelet for FWI, depicted as the red line in Figure \ref{fig:wavelet_spectra}a. The spectra of both the true and erroneous wavelets are shown in Figure \ref{fig:wavelet_spectra}b, clearly indicating differences in their frequency components.

\begin{figure}
\begin{center}
\includegraphics[width=1.0\textwidth]{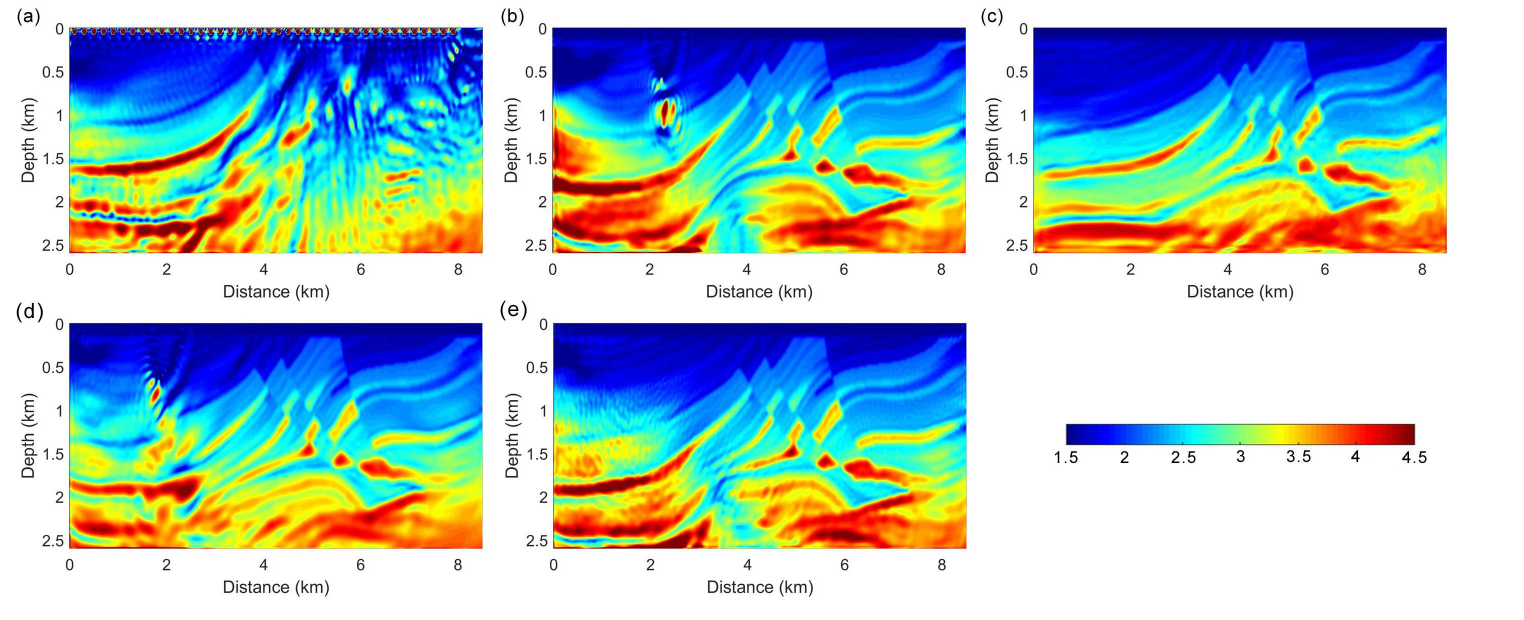} 
\caption{The inverted velocity models from (a) FWI with the wrong wavelet, (b) FWI with the true wavelet, (c) the proposed DL-SIFWI with the wrong wavelet, (d) envelope-based FWI with the true wavelet, and (e) Wasserstein-distance-based FWI with the true wavelet.}
\label{fig:mar_inv}
\end{center}   
\end{figure}

Initially, we conduct the conventional FWI using the wrong wavelet for 2000 iterations. The resulting inverted velocity model is shown in Figure \ref{fig:mar_inv}a. It is evident that the model suffers from significant source footprint artifacts due to the error in the wavelet, with no detailed structures recovered. Next, we perform the conventional FWI using the true wavelet, resulting in the model displayed in Figure \ref{fig:mar_inv}b. While it mitigates the source footprint errors in the shallow region, it still fails to adequately reconstruct the true model because the observed data lacks low frequencies.

Subsequently, we apply the proposed DL-SIFWI method with the wrong wavelet. The size of the OpenFWI velocity is generally small with $70 \times 70$. We stack 40 velocities as the input. Thus, the size of the input to the framework of DI-SIFWI is $40 \times 70 \times 70$. This configuration is applied to all the experiments. During the 2000 epochs of training, we adjust the weight factor $\alpha$ in the DL-SIFWI objective function as specified in equation \ref{eqn:eq5}. The values of $\alpha$ for different training epochs are summarized in Table 1, where we gradually decrease $\alpha$ to enhance the data-fitting term. The inverted velocity model obtained via DL-SIFWI is shown in Figure \ref{fig:mar_inv}c, which accurately recovers the structures of the true model. We further compare the DL-SIFWI-inverted velocity model with inverted velocities obtained from two advanced FWI approaches: envelope-based FWI and optimal transport-based FWI. Using the same inversion setup and the true source wavelet, we first apply weighted envelope inversion to recover the smooth background velocity, followed by standard FWI to enhance the resolution of the velocity model, as described by \cite{song2023weighted}. The resulting inverted velocity model is shown in Figure \ref{fig:mar_inv}d. Although the envelope-based FWI achieves better recovery of the deep structure compared to the conventional FWI, it still exhibits significant velocity errors in the left part of the model, primarily due to cycle skipping. Similarly, we implement Wasserstein-distance-based FWI, leveraging the theory of optimal transport, to obtain the inverted velocity model displayed in Figure \ref{fig:mar_inv}e. While this method improves the shallow structure, the deeper regions continue to suffer from cycle skipping.

\begin{figure}
\begin{center}
\includegraphics[width=1.0\textwidth]{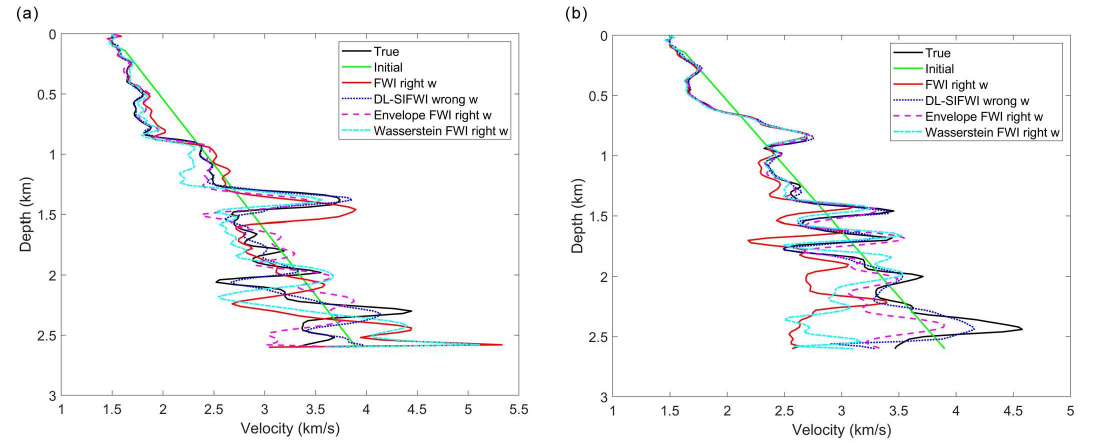} 
\caption{Comparison of vertical velocity profiles at locations (a) 2 km and (b) 4 km.}
\label{fig:mar_profiles}
\end{center}   
\end{figure}

To facilitate a detailed comparison of the inverted velocities, we present vertical velocity profiles at locations 2 km and 4 km in Figures \ref{fig:mar_profiles}a and \ref{fig:mar_profiles}b, respectively. These profiles illustrate that the inverted velocities (red solid curve) from the conventional FWI significantly deviate from the true model. While envelope-based FWI (magenta dashed curve) and Wasserstein-distance-based FWI (cyan dotted-dashed curve) provide some improvement over conventional FWI, neither aligns well with the true velocity in the deeper regions. In contrast, the DL-SIFWI-inverted velocity (blue dotted curve) closely matches the true velocity (solid black curve) even with the wrong wavelet, indicating its superior accuracy. 

\begin{table}[!htbp]
\centering
\caption{The weight factor $\alpha$ for different training epochs.}\label{tab:aStrangeTable}%
\begin{tabular}{cc}
\toprule
Epoch range & $\alpha$ \\
\midrule
1-500 & $10^{-7}$\\
501-1000 & $5 \times 10^{-8}$\\
1001-1500 & $1 \times 10^{-8}$\\
1501-2000 & 0.0 \\
\bottomrule
\end{tabular}
\end{table} 

\begin{figure}
\begin{center}
\includegraphics[width=0.5\textwidth]{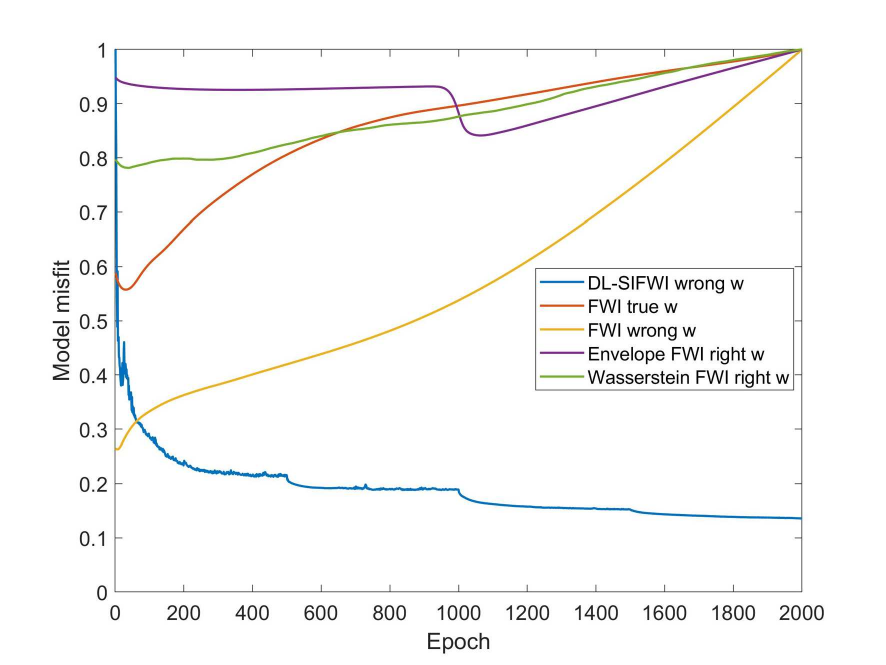} 
\caption{The model misfit curves from different FWI strategies. (Blue line: DL-SIFWI; red line: FWI with the true wavelet; yellow line: FWI with the wrong wavelet; purple line: envelope FWI; green line: Wasserstein FWI.)}
\label{fig:mar_loss}
\end{center}   
\end{figure}

The normalized model misfit curves for the five implemented FWI strategies are illustrated in Figure \ref{fig:mar_loss}. We observe that the model misfits increase for the conventional FWI, both with and without the true wavelet, due to the source wavelet error and the cycle-skipping issue. The model misfit curves from envelope-based FWI and Wasserstein-distance-based FWI also show an increase, albeit to a lesser extent, indicating that while the cycle-skipping issue is mildly reduced, it is not entirely resolved. In contrast, the model misfit for DL-SIFWI consistently decreases until convergence. This result came in spite of missing low frequency, starting with a poor initial velocity and using an inaccurate wavelet for modeling.

\begin{figure}
\begin{center}
\includegraphics[width=1.0\textwidth]{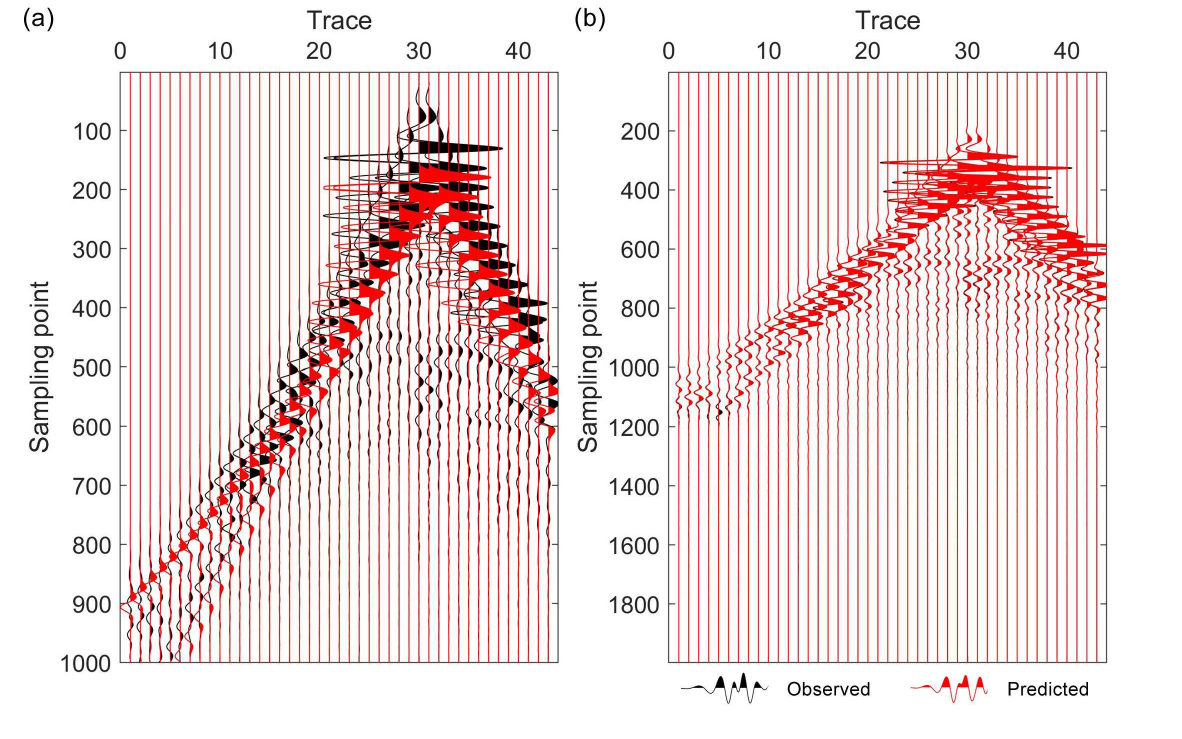} 
\caption{(a) Comparison between shot gathers from the observed and predicted data using the wrong wavelet and initial velocity. (b) The comparison between convolved shot gathers from the observed and predicted data using the wrong wavelet and inverted velocity.)}
\label{fig:mar_data_comparison}
\end{center}   
\end{figure}

Finally, we compare the data fitting, which is a primary objective of FWI before and after implementing the proposed DL-SIFWI. Figure \ref{fig:mar_data_comparison}a shows the comparison between shot gathers from the observed data (marked in black) and the predicted data (marked in red) corresponding to the initial velocity with the wrong wavelet. It is evident that they do not match, even for the direct waves, due to a significant time shift in the source wavelets used. In contrast, the convolved observed data (black) and the convolved predicted data (red), utilizing the wrong wavelet and the inverted velocity, almost completely overlap (Figure \ref{fig:mar_data_comparison}b). This demonstrates the accuracy of the inverted velocity. We find that even a wrong wavelet, the predicted data can still effectively match the observed data after reference trace convolution. For all the examples implemented in this paper, we select the reference trace towerd the near offset as it contains the most source information \cite{song2020source}. It is worth noting that the sampling points on the vertical axis in Figure \ref{fig:mar_data_comparison}b are nearly double those in Figure \ref{fig:mar_data_comparison}a due to the convolution operation.

In this example, we train the DL-SIFWI network for 2000 epochs to obtain the inverted velocity model, which takes 51.8 minutes to complete. All experiments in this paper were conducted on an RTX A6000 GPU. For comparison, the conventional FWI with 2000 iterations requires 49.7 minutes to run on the same hardware. This suggests that the computational demand of DL-SIFWI is generally affordable in terms of time and the DL-SIFWI method has the potential for improved model quality, especially when dealing with complex subsurface structures and a wrong source wavelet.

\subsection{Overthrust model}

\begin{figure}
\begin{center}
\includegraphics[width=1.0\textwidth]{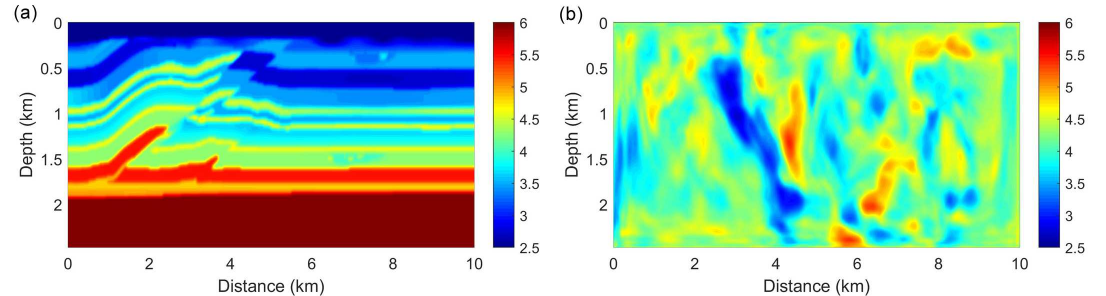} 
\caption{(a) The true Overthrust model, and (b) the randomly initialized velocity model.}
\label{fig:over_true_ini}
\end{center}   
\end{figure}

Then, we test the proposed DL-SIFWI method on the Overthrust model, which has a grid size of $401 \times 100$ and a vertical and horizontal sampling interval of 25 m. The true velocity model is shown in Figure \ref{fig:over_true_ini}a. To further evaluate the capability of the proposed method in mitigating cycle skipping, we do not provide any initial velocity as prior information, setting $\alpha=0$ in equation \ref{eqn:eq5}. The randomly initialized velocity model generated by the autoencoder, displayed in Figure \ref{fig:over_true_ini}b, contains no information about the true model, other than it is constrained between the set minimum and maximum velocity guaranteed by the sigmoid activation function. We again use a 6-Hz Ricker wavelet (blue line in Figure \ref{fig:wavelet_spectra}a) to generate the observed data without frequencies below 3 Hz , with 40 shots evenly distributed on the surface. All grid points at a depth of 25 m serve as receivers to record the observed data.

\begin{figure}
\begin{center}
\includegraphics[width=1.0\textwidth]{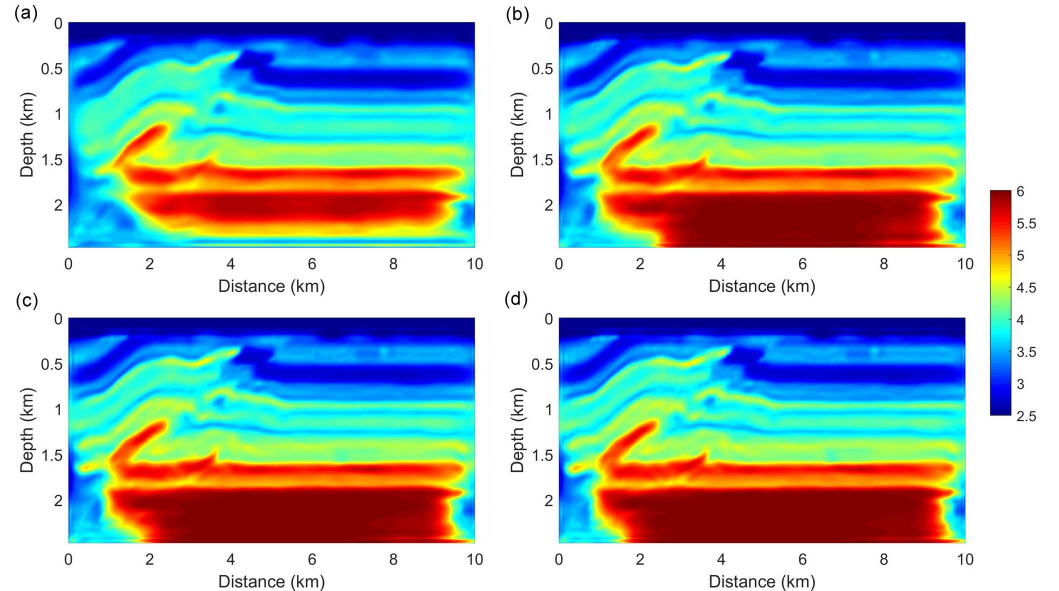} 
\caption{The DL-SIFWI-inverted velocity models after (a) 500, (b) 1000, (c) 1500, and (d) 2000 epochs with observed data without low frequencies below 3 Hz.}
\label{fig:over_3hz_inv}
\end{center}   
\end{figure}

We utilize the same wrong wavelet (red line in Figure \ref{fig:wavelet_spectra}b) for the DL-SIFWI. Despite the extremely poor initial condition, lacking low-frequency data, an initial velocity, and an accurate wavelet, DL-SIFWI successfully recovers the structures of the true model. The inverted velocity models after 500, 1000, 1500, and 2000 epochs are shown in Figures \ref{fig:over_3hz_inv}a-\ref{fig:over_3hz_inv}d. The shallow part of the Overthrust model is well recovered after 500 epochs. As the training progresses, the high-velocity base in the deeper area is gradually reconstructed. The final inverted velocity exhibits only minor errors in the deep corners due to the illumination limitation.

\begin{figure}
\begin{center}
\includegraphics[width=1.0\textwidth]{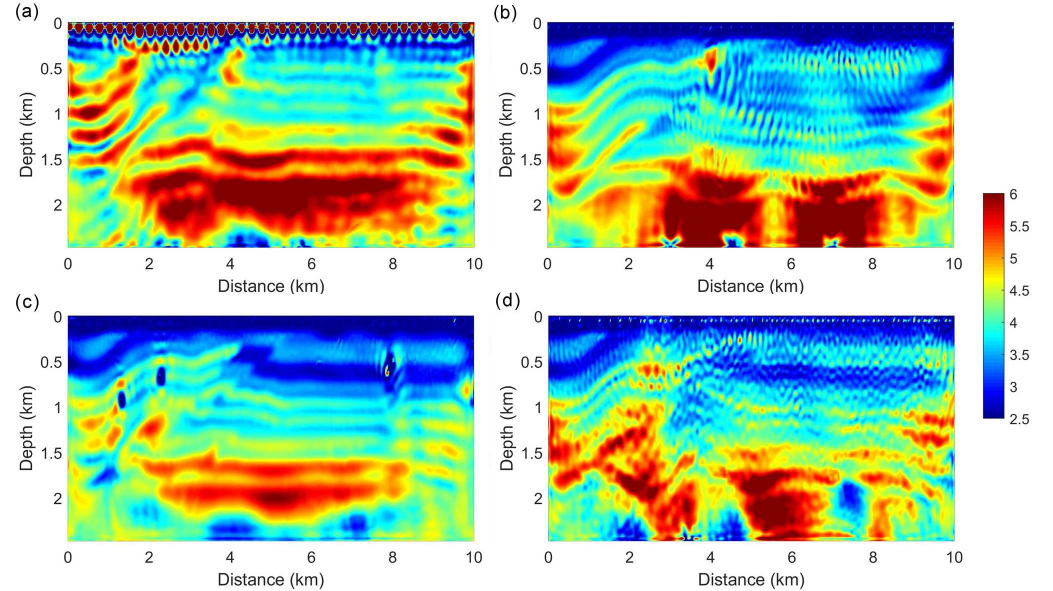} 
\caption{(a) The FWI-inverted velocity models with the (a) wrong, (b) true source wavelets, (c) envelope-based FWI with the true wavelet, and (d) Wasserstein-distance-based FWI with the true wavelet.}
\label{fig:over_fwi}
\end{center}   
\end{figure}

We perform conventional FWI with the same number of iterations, starting with the velocity shown in Figure \ref{fig:over_true_ini}b, using both the wrong and true source wavelets. The results are displayed in Figures \ref{fig:over_fwi}a and \ref{fig:over_fwi}b, respectively. Neither FWI-inverted velocity model resembles the true model. While the true wavelet helps in the shallow area, deep structures remain severely distorted due to cycle skipping caused by the crude initial velocity in Figure \ref{fig:over_true_ini}b and the absence of low-frequency data. Using the true source wavelet, the weighted envelope-based FWI successfully reconstructs the shallow structures with only minor errors but struggles to accurately recover the deep structures due to the poor quality of the initial velocity model, as shown in Figure \ref{fig:over_fwi}c. Similarly, Figure \ref{fig:over_fwi}d demonstrates that the Wasserstein-distance-based FWI achieves slight improvements over conventional FWI in reconstructing the shallow structures.

\begin{figure}
\begin{center}
\includegraphics[width=1.0\textwidth]{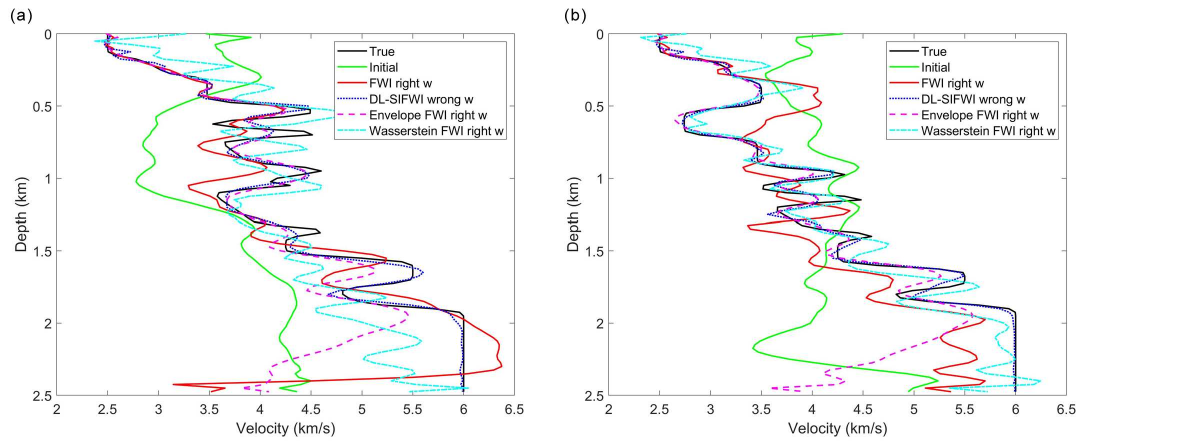} 
\caption{Comparison of vertical velocity profiles at the locations (a) 3 km and (b) 6 km.}
\label{fig:over_profiles}
\end{center}   
\end{figure}

We compare the vertical velocity profiles at locations 3 km and 6 km, as shown in Figures \ref{fig:over_profiles}a and \ref{fig:over_profiles}b, respectively. The velocity profiles inverted by DL-SIFWI closely match the true ones, whereas the velocities from the conventional FWI diverge significantly. The velocity profiles obtained from envelope-based FWI and Wasserstein-distance-based FWI using the true wavelet show poor agreement with the true velocity, particularly in the deep regions.

\begin{figure}
\begin{center}
\includegraphics[width=1.0\textwidth]{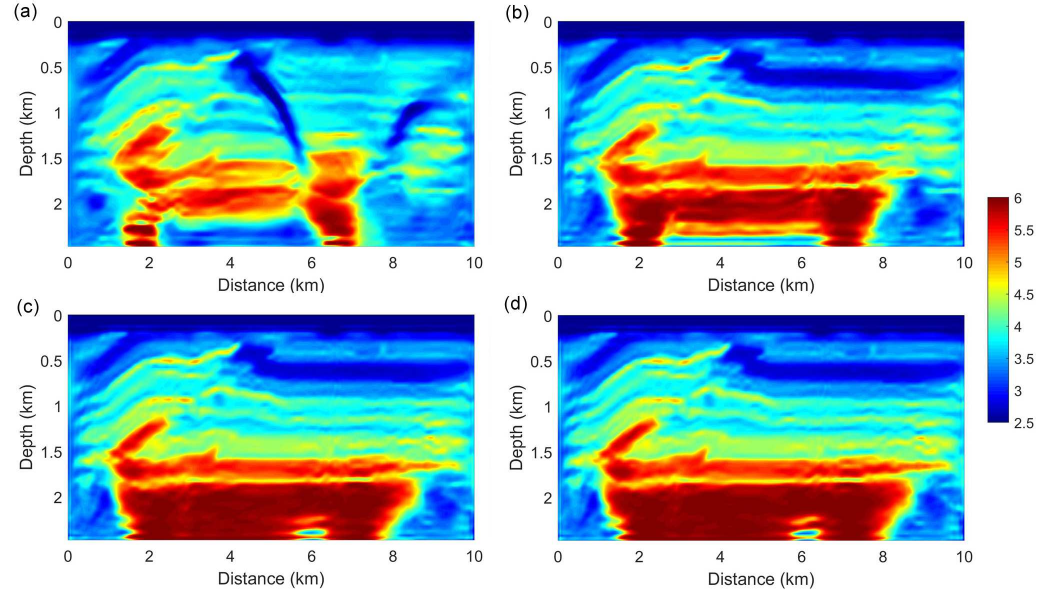} 
\caption{The DL-SIFWI-inverted velocity models after (a) 1000, (b) 2000, (c) 3000, and (d) 4000 epochs with observed data without low frequencies below 5 Hz.}
\label{fig:over_5hz_inv}
\end{center}   
\end{figure}

To further challenge the proposed method, we generate observed data using a new wavelet with a dominant frequency of 8 Hz, muting low frequencies below 5 Hz. We continue using the same wrong wavelet for DL-SIFWI. This scenario is considered very difficult for conventional FWI due to inevitable cycle skipping and errors in the source wavelet. Nevertheless, DL-SIFWI successfully predicts a reasonable velocity by optimizing the network under the constraint of data fitting. The DL-SIFWI-inverted velocity models after 1000, 2000, 3000, and 4000 epochs, using observed data without low frequencies below 5 Hz, are shown in Figures \ref{fig:over_5hz_inv}a-\ref{fig:over_5hz_inv}d, respectively. Though the inverted velocity after 1000 epochs contains noticeable errors from cycle skipping (Figure \ref{fig:over_5hz_inv}a), the DL-SIFWI method gradually corrects these errors with increasing training epochs, as illustrated in Figures \ref{fig:over_5hz_inv}b-\ref{fig:over_5hz_inv}d thanks to the extended model space given by the network parameters.

\subsection{Real marine data}

\begin{figure}
\begin{center}
\includegraphics[width=1.0\textwidth]{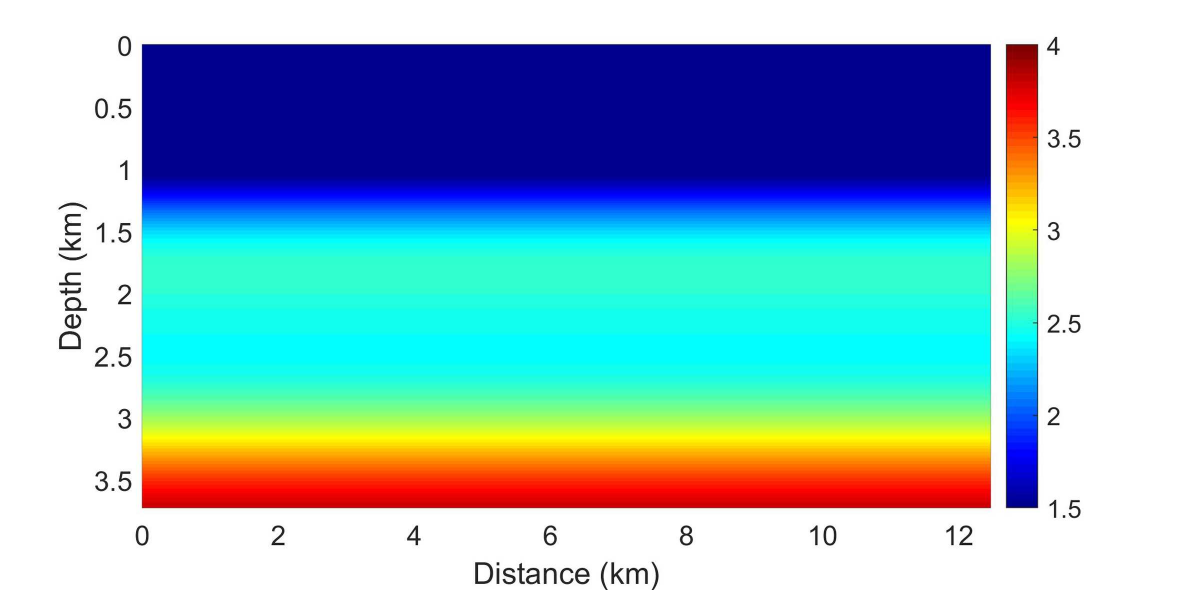} 
\caption{The 1D initial velocity model.}
\label{fig:cgg_v0_1d}
\end{center}   
\end{figure}

Next, we show the effectiveness of DL-SIFWI on a real dataset acquired by Viridien (formerly CGG) from the North Western Australia continental shelf \cite{soubaras2010variable}. We selected 100 shot gathers from the original data, covering a target area that spans 12.5 km in distance and 3.75 km in depth. The corresponding shots we selected are evenly distributed within the range of 0 km to 9.28 km. We use a 1D velocity that varies only vertically as the initial velocity model, as shown in Figure \ref{fig:cgg_v0_1d}. The vertical and horizontal spatial sampling intervals of the model are both set to 25 m.

\begin{figure}
\begin{center}
\includegraphics[width=1.0\textwidth]{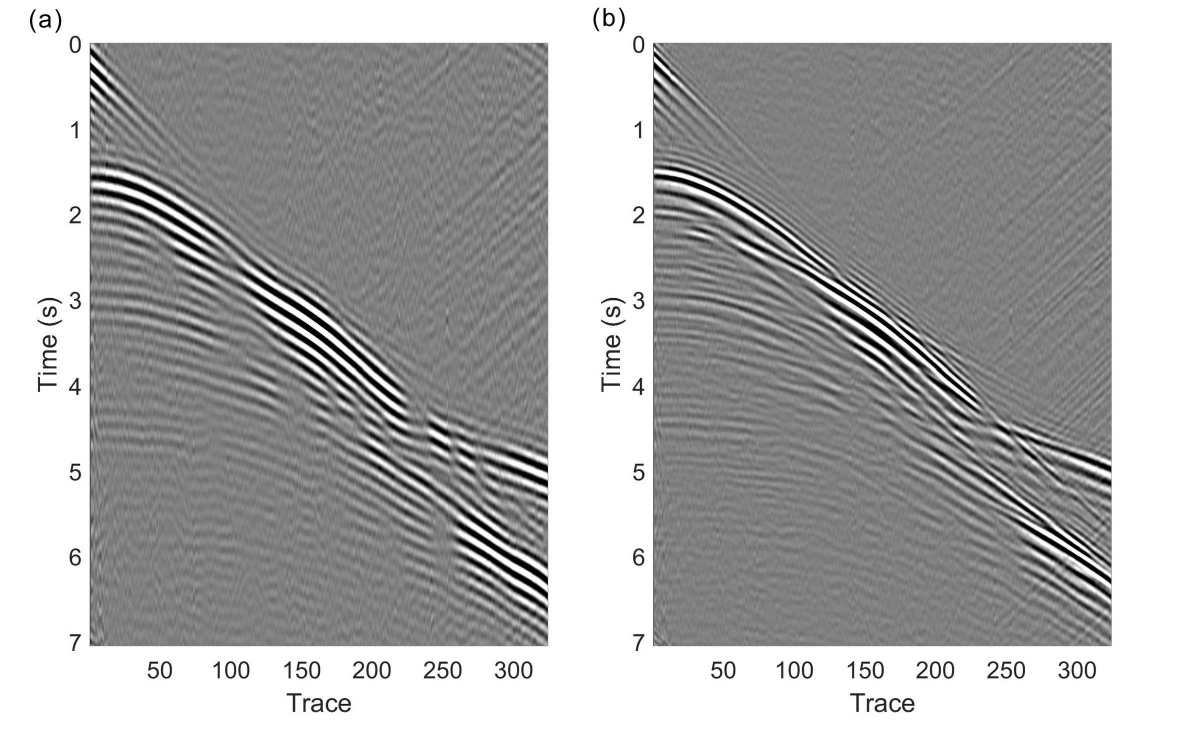} 
\caption{An example filtered shot gather with frequency bands of (a) 2-8 Hz and (b) 2-12 Hz.}
\label{fig:cgg_data}
\end{center}   
\end{figure}

\begin{figure}
\begin{center}
\includegraphics[width=0.5\textwidth]{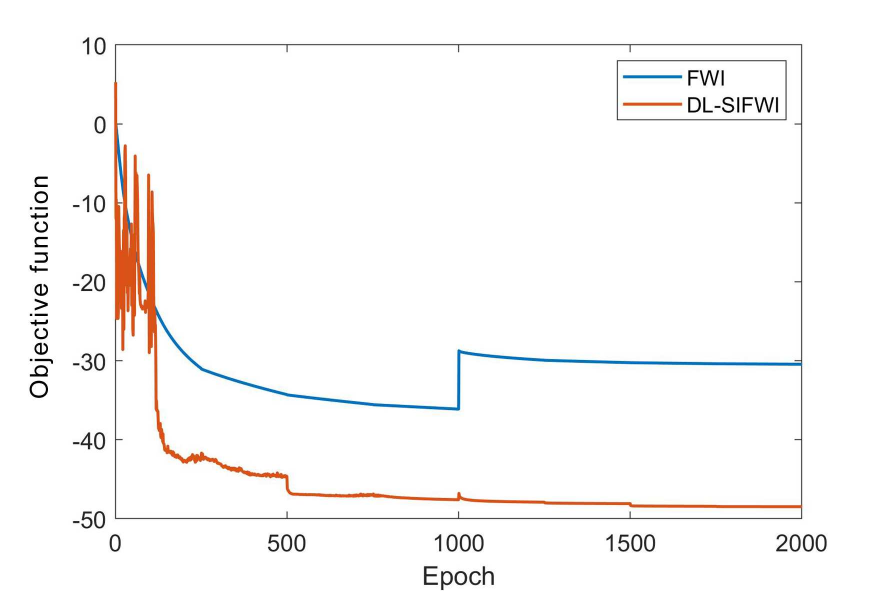} 
\caption{The loss curves of the objective functions. (Blue line: DL-SIFWI; red line: FWI.)}
\label{fig:cgg_data_loss}
\end{center}   
\end{figure}

\begin{figure}
\begin{center}
\includegraphics[width=1.0\textwidth]{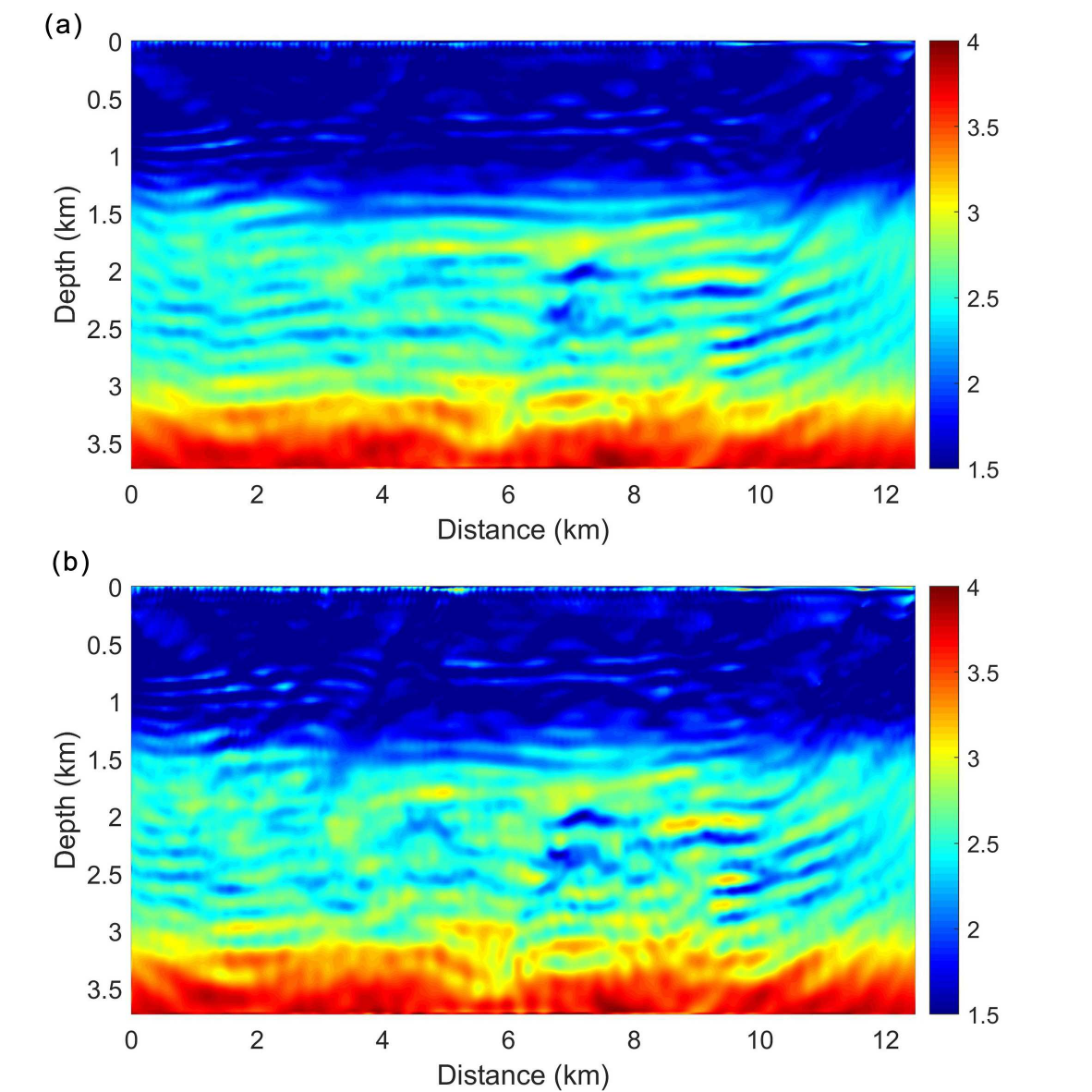} 
\caption{The FWI-inverted velocity models using (a) 2-8 Hz and (b) 2-12 Hz filtered data.}
\label{fig:cgg_fwi}
\end{center}   
\end{figure}

We conduct multi-stage FWI and DL-SIFWI using two sets of bandpass filtered data: 2-8 Hz and 2-12 Hz. One filtered shot gather for these frequency bands are displayed in Figures \ref{fig:cgg_data}a and \ref{fig:cgg_data}b, respectively. We do not estimate the source wavelet from the data and instead still use the 6-Hz Ricker wavelet (blue line in Figure \ref{fig:wavelet_spectra}a) for all sources during both FWI and DL-SIFWI implementations. For both inversion strategies, we begin with the low-frequency band (2-8 Hz) to ensure a robust update of the background velocity. Sequentially, we employ the 2-12 Hz frequency band to incorporate higher frequencies for refining the inverted velocity resolution. Each inversion stage involve 1000 epochs to update the velocity. For the FWI implementation, we utilize the GCN objective function, known for its adaptability to field data \cite{wu2017efficient}. The loss curve of FWI across different epochs is illustrated in Figure \ref{fig:cgg_data_loss} as the blue curve. Notably, there is a sudden increase in the loss curve when transitioning to the high-frequency band, reflecting the challenge in data fitting at higher frequencies. Although the overall trend of the FWI loss curve shows a gradual decrease, that decrease can be to a local minimum with respect to a portion of the model parameters, especially those representing the deeper part of the model. This is mainly due to the crude initial velocity used here. The FWI-inverted velocity models for the 2-8 Hz and 2-12 Hz filtered data are shown in Figures \ref{fig:cgg_fwi}a and \ref{fig:cgg_fwi}b, respectively. The results indicate that FWI struggles to capture continuous sedimentary layers or reflect any reasonable geological features.

\begin{figure}
\begin{center}
\includegraphics[width=1.0\textwidth]{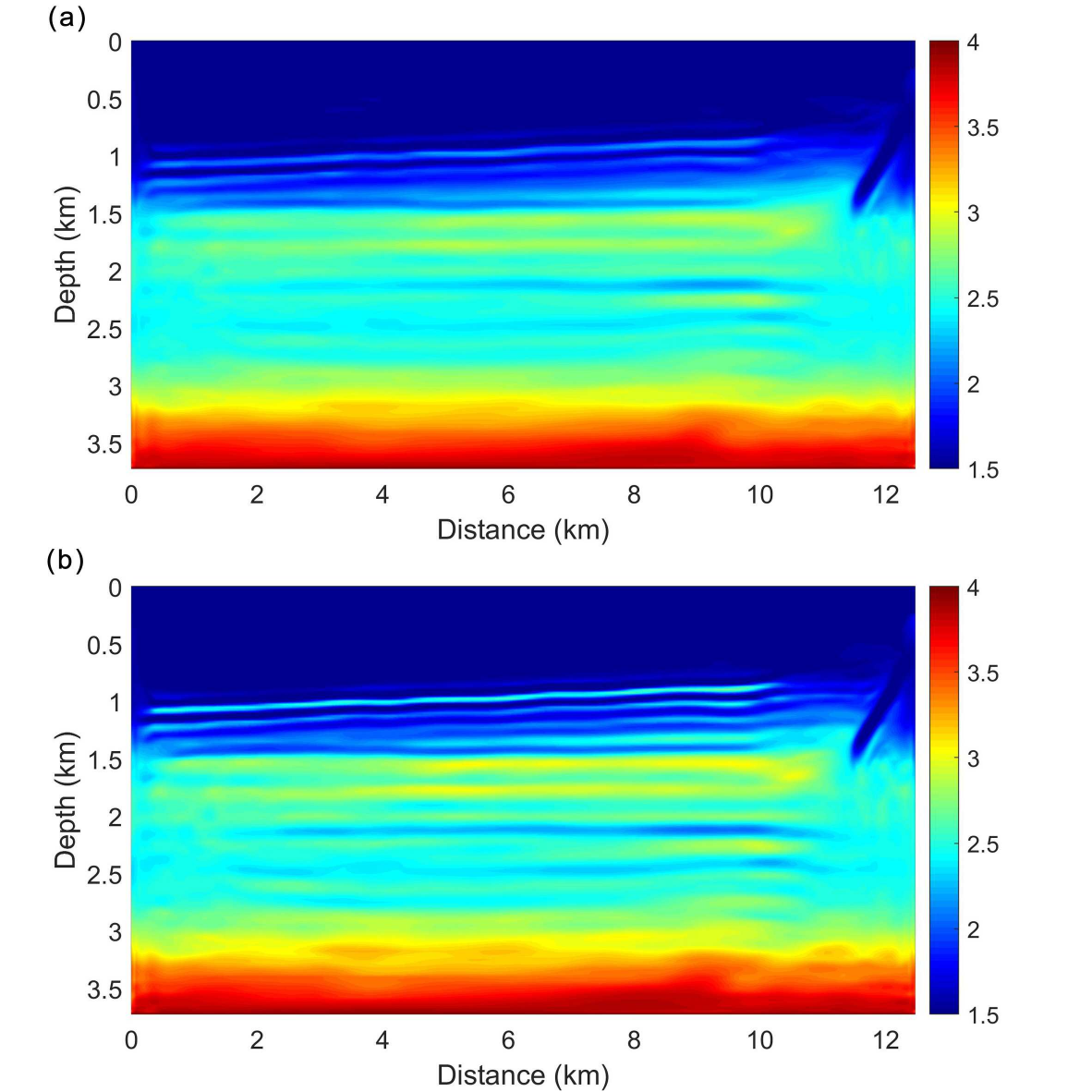} 
\caption{The DL-SIFWI-inverted velocity models using (a) 2-8 Hz and (b) 2-12 Hz filtered data.}
\label{fig:cgg_dlsifwi}
\end{center}   
\end{figure}

In the DL-SIFWI approach, we set $\alpha$ to $ 2 \times 10^{-6}$ for the first 500 epochs and halve it for the next 500 epochs to enhance data fitting during both inversion stages. Although the loss curve of DL-SIFWI exhibits significant oscillations in the beginning, it decreases rapidly and achieves reasonable convergence, as shown by the red curve in Figure \ref{fig:cgg_data_loss}. Figures \ref{fig:cgg_dlsifwi}a and \ref{fig:cgg_dlsifwi}b display the DL-SIFWI-inverted velocity models for the 2-8 Hz and 2-12 Hz filtered data, respectively. These models successfully capture high-velocity layers between 1.5-2.0 km depth and low-velocity layers between 2.0-2.5 km depth.

\begin{figure}
\begin{center}
\includegraphics[width=1.0\textwidth]{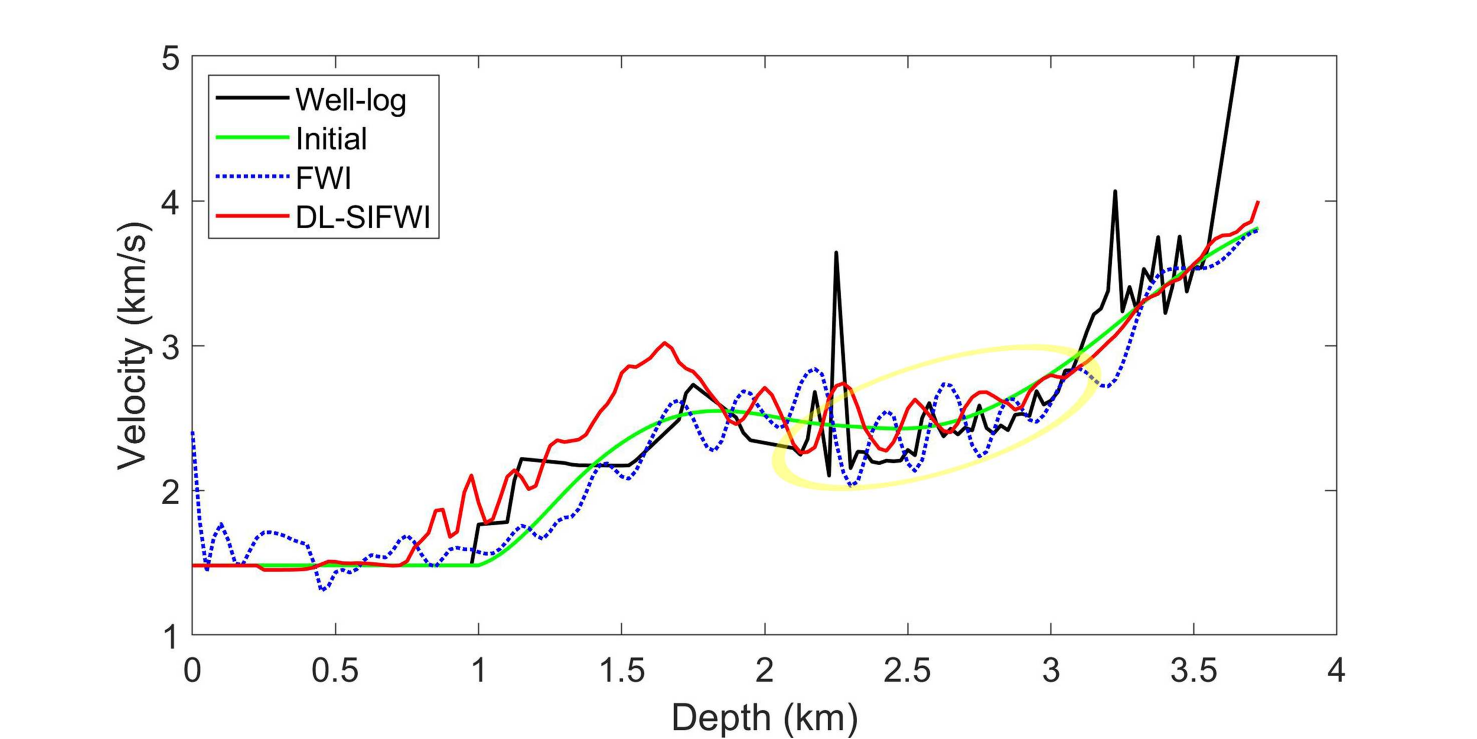} 
\caption{Comparison of the vertical velocity profiles at location of 10.5 km. The yellow circle indicates a reasonable fitting between the DL-SIFWI-inverted velocity (red solid line) and the check-shot velocity (black solid line).}
\label{fig:cgg_profile}
\end{center}   
\end{figure}

\begin{figure}
\begin{center}
\includegraphics[width=1.0\textwidth]{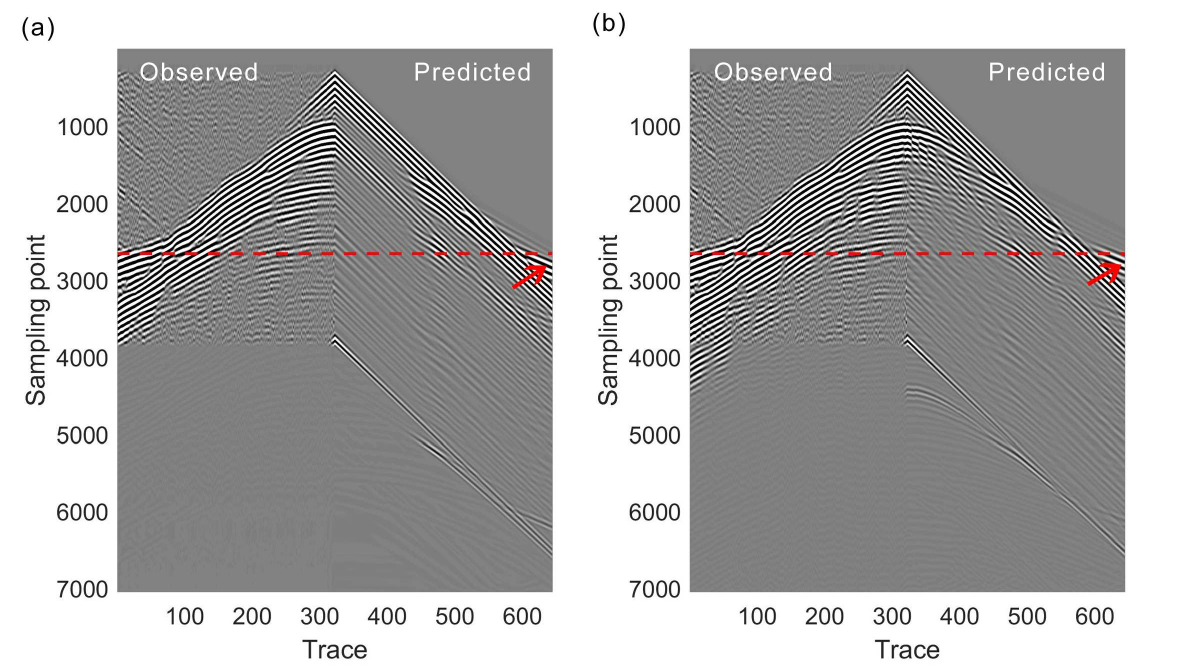} 
\caption{A comparison between the convolved observed data with the convolved predicted data from the (a) initial and (b) DL-SIFWI-inverted velocity models.}
\label{fig:cgg_data_comparison}
\end{center}   
\end{figure}

To verify the accuracy of the DL-SIFWI-inverted velocity model, we compare it with a well-log velocity profile obtained from a check-shot measurement at location 10.5 km. The comparison is illustrated in Figure \ref{fig:cgg_profile}, where we observe a reasonable agreement between the DL-SIFWI-inverted velocity and the well-log velocity, particularly in the circled area. In contrast, the vertical velocity profile from FWI (blue dotted curve) does not match any layers in the well-log velocity . Additionally, we present comparisons between the convolved observed data and the convolved predicted data from the initial and inverted velocity models in Figures \ref{fig:cgg_data_comparison}a and \ref{fig:cgg_data_comparison}b, respectively. In Figure \ref{fig:cgg_data_comparison}a, the convolved predicted data do not match with the reflections in the convolved observed data, and a noticeable time shift occurs in the refractions at the far offset, as indicated by the red dashed line and arrow. In contrast, the convolved predicted data from the inverted velocity model matches well with the convolved observed data in both reflections and refractions (Figure \ref{fig:cgg_data_comparison}b), confirming the accuracy of the inverted model.

\subsection{Real OBC data}

\begin{figure}
\begin{center}
\includegraphics[width=0.6\textwidth]{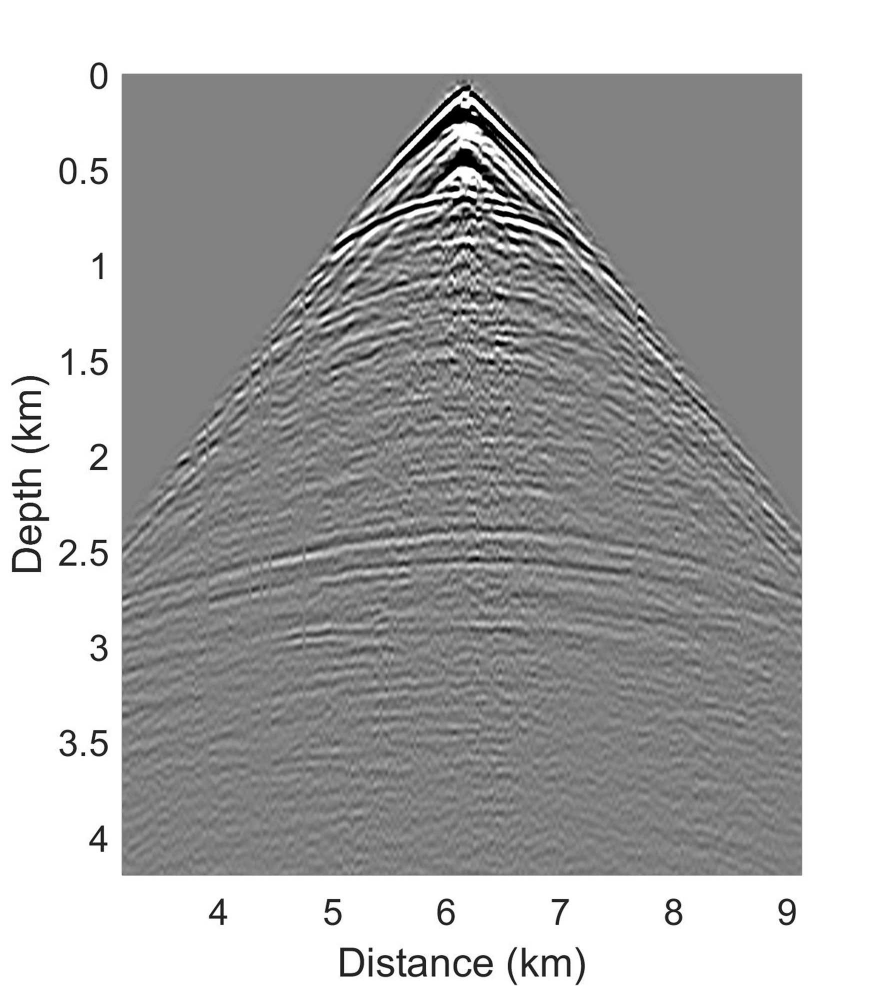} 
\caption{One vertical-component shot gather.}
\label{fig:volve_data61}
\end{center}   
\end{figure}

In the final test, we apply the proposed method to a 2D section extracted from a 3D Volve ocean bottom cable (OBC) dataset acquired in the North Sea \cite{szydlik20103d}. This 2D section includes 121 shots spaced 100 m apart, with each shot recorded by 240 receivers at a 25 m interval. Both vertical and horizontal components of the data are provided after processing. A vertical-component shot gather is displayed in Figure \ref{fig:volve_data61}, where we observe that the direct and diving waves have been muted, potentially posing challenges for FWI.

\begin{figure}
\begin{center}
\includegraphics[width=1.0\textwidth]{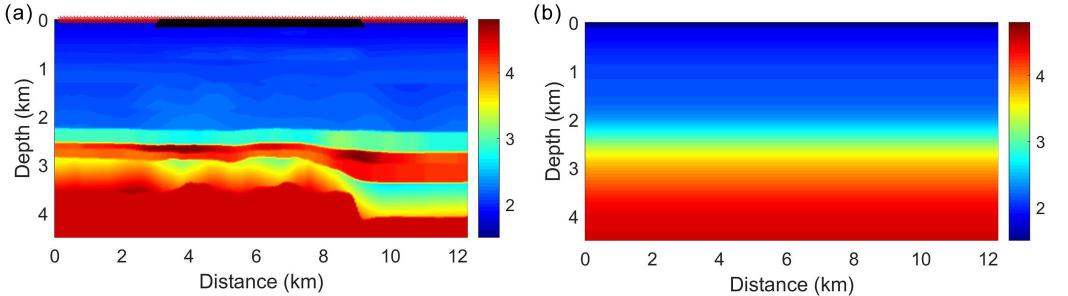} 
\caption{(a) A legacy velocity model obtained from tomography.(Red $*$: sources locations; black $\bigtriangleup$: receivers locations.) (b) The 1D initial velocity model.}
\label{fig:volve_tomo_ini}
\end{center}   
\end{figure}

Figure \ref{fig:volve_tomo_ini}a shows the legacy P-wave velocity model obtained through tomography, covering a survey area of 12.3 km $\times$ 4.25 km. The red $*$ and black $\bigtriangleup$ markers denote the source and receiver locations, respectively. The model reveals a high-velocity seal layer at a depth of approximately 2.75 km and a low-velocity region beneath it, interpreted as a reservoir. A 1D velocity model without horizontal variation, shown in Figure \ref{fig:volve_tomo_ini}b, was used as the initial velocity model. This initial model lacks any indication of the seal layer or reservoir features, providing a challenging starting point for the inversion process.

\begin{figure}
\begin{center}
\includegraphics[width=1.0\textwidth]{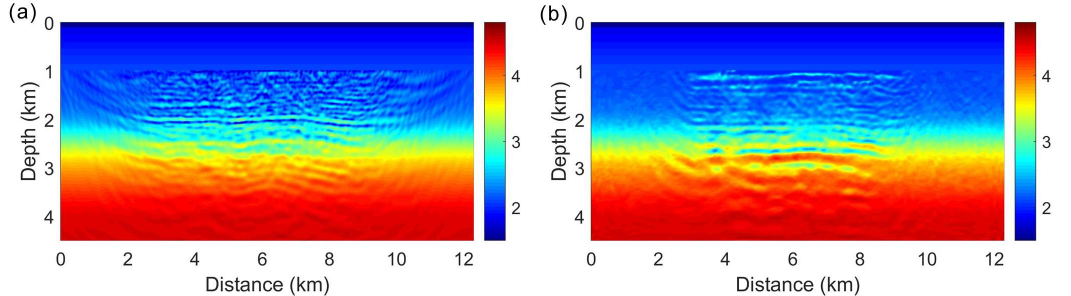} 
\caption{The (a) FWI-inverted and (b) DL-SIFWI-inverted velocity models .}
\label{fig:volve_tomo_fwi_dlsifwi}
\end{center}   
\end{figure}

In this test, we directly utilized the processed vertical-component data to invert the P-wave velocity without employing a multi-scale strategy that starts from the low-frequency band. A 6-Hz Ricker wavelet was used as the source function for the inversion. The velocity above 1 km was not updated due to the absence of diving waves, which are typically essential for resolving shallow structures. After 1000 iterations, the FWI-inverted velocity model is shown in Figure \ref{fig:volve_tomo_fwi_dlsifwi}a. It is evident that FWI fails to resolve the seal layer and reservoir within the survey area. With the same number of iterations, DL-SIFWI was employed with a varying weight factor $\alpha$ across different training epochs (as detailed in Table 2). The resulting velocity model, shown in Figure \ref{fig:volve_tomo_fwi_dlsifwi}b, demonstrates that DL-SIFWI successfully reconstructs the key structural features of the seal layer and reservoir, highlighting its effectiveness in this challenging scenario. 

\begin{table}[!htbp]
\centering
\caption{The weight factor $\alpha$ for different training epochs in the Volve data inversion.}
\label{tab:aStrangeTable2}%
\begin{tabular}{cc}
\toprule
Epoch range & $\alpha$ \\
\midrule
1-250 & $4 \times10^{-7}$\\
251-500 & $3 \times 10^{-7}$\\
501-750 & $2 \times 10^{-7}$\\
751-1000 & $1 \times 10^{-7}$ \\
\bottomrule
\end{tabular}
\end{table} 

\begin{figure}
\begin{center}
\includegraphics[width=0.6\textwidth]{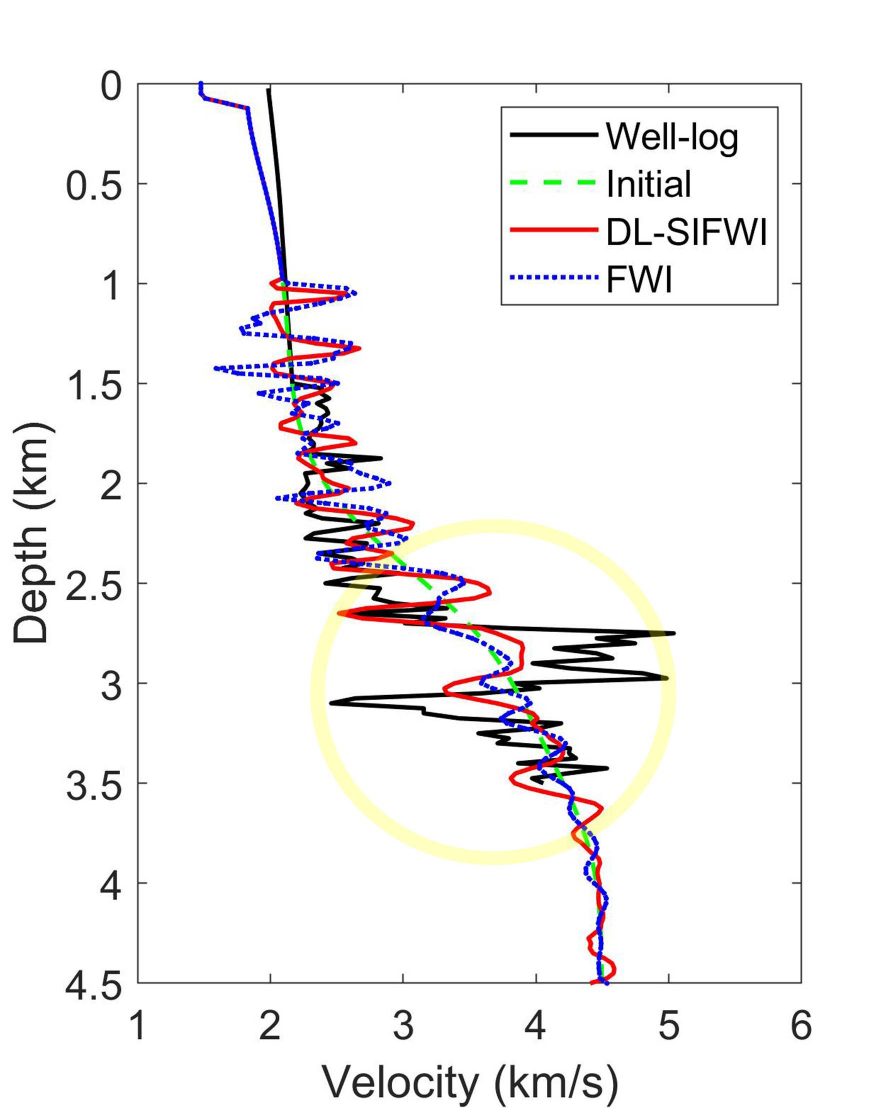} 
\caption{Comparison of the vertical velocity profiles at location of 7.75 km. The yellow circle indicates a reasonable fitting between the DL-SIFWI-inverted velocity (red solid line) and the check-shot velocity (black solid line).}
\label{fig:volve_profile}
\end{center}   
\end{figure}

\begin{figure}
\begin{center}
\includegraphics[width=1.0\textwidth]{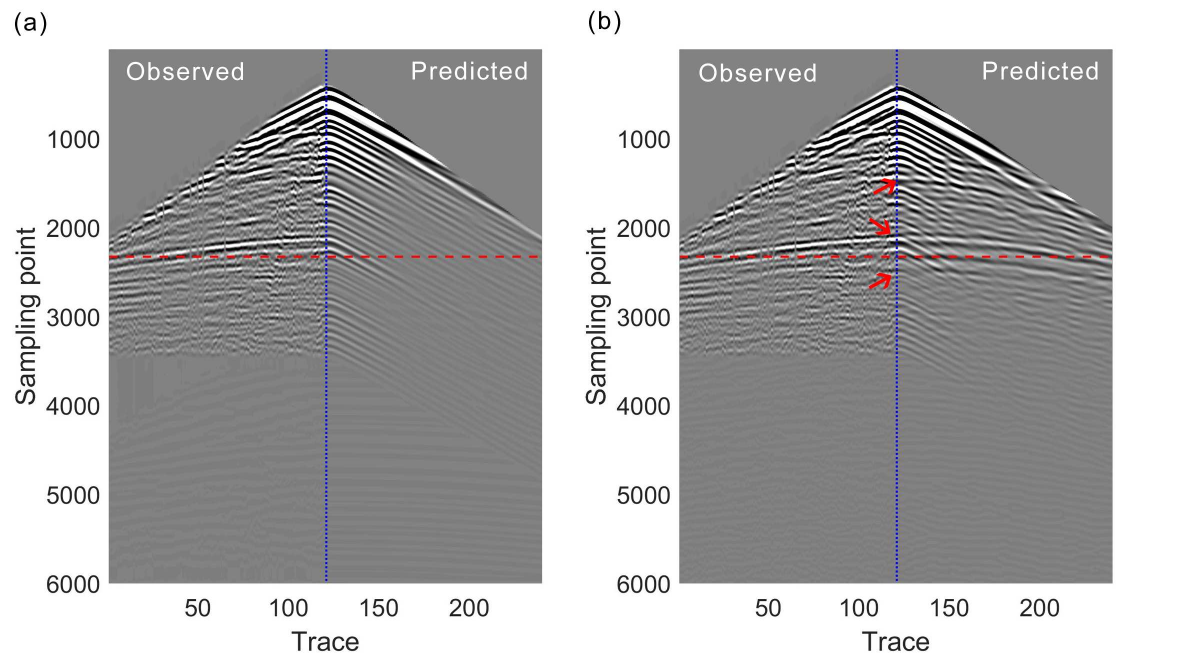} 
\caption{A comparison between the convolved observed data with the convolved predicted data from the (a) initial and (b) DL-SIFWI-inverted velocity models.} 
\label{fig:volve_data_comparison}
\end{center}   
\end{figure}

We validate the accuracy of the inverted velocity models by comparing them with an available check-shot velocity profile at 7.75 km. The comparison, presented in Figure \ref{fig:volve_profile}, demonstrates that the DL-SIFWI-inverted velocity (red solid line) aligns closely with the check-shot velocity profile (black solid line), particularly in the circled area. In contrast, the FWI-inverted velocity (blue dotted line) exhibits noticeable discrepancies. Furthermore, we compare the convolved data fitting between the observed data and the predicted data generated from the initial and DL-SIFWI-inverted velocity models, as shown in Figures \ref{fig:volve_data_comparison}a and \ref{fig:volve_data_comparison}b, respectively. Figure \ref{fig:volve_data_comparison}a indicates that the predicted convolved data from the initial velocity model poorly matches the reflections in the observed convolved data. By comparison, as shown in Figure \ref{fig:volve_data_comparison}b, the reflections in the observed convolved data are well-aligned with those in the predicted convolved data from the DL-SIFWI-inverted velocity model at both near and far offsets, as highlighted by the red arrows and dashed line.

\section{Discussion}

In conventional FWI, the target parameters in the optimization process are the velocity values at the meshing grids. Because the forward modeling operator (wave equation) is highly dependent on the velocity, FWI is inherently nonlinear, making it prone to cycle skipping. In contrast, DL-based FWI reparameterizes the velocity model using the trainable parameters of deep networks. The number of trainable parameters in DL-based FWI is typically much larger than the number of grid points representing the velocity model. This introduces a significantly larger number of degrees of freedom or search space into the inverse problem, which helps to overcome local minima. Moreover, DL-based FWI exhibits less reliance on low-frequency seismic data. Neural networks inherently favor smooth representations of the velocity model due to spectral bias during the early stages of the inversion process. As the inversion progresses, neural networks perform hierarchical learning to progressively add finer details to the velocity model.

The resolution of the resulting inverted velocity model in DL-SIFWI is primarily determined by the architecture of the autoencoder network. Larger and deeper autoencoder networks can capture more complex features and finer details in the velocity model, leading to higher-resolution inversions. This is because deeper networks with more layers and parameters are able to learn more intricate representations of underlying geological structures. In contrast, smaller networks with fewer layers and parameters may struggle to resolve high-frequency components or finer details, resulting in lower-resolution outputs. However, from the efficiency perspective, larger networks come with an increased computational cost for training. In this study, we use six encoder blocks and six decoder blocks for all experiments. This configuration was chosen based on intuition, without extensive trial-and-error testing, and yielded reasonably good results. We used 2000 epochs to train the proposed DL-SIFWI framework in all the examples. Though the number is relatively high, the computational time is acceptable. For the Marmousi model, it takes less than one hour to finish the inversion process using one GPU. The quantitative relationship between network size and velocity resolution remains an open question. Understanding this relationship will provide valuable insights into selecting an appropriate network architecture fitting the resolution requirements of specific seismic inversion tasks. We will conduct further research to explore this topic.

The training strategy is fully unsupervised, focusing solely on the core objective of FWI: data fitting. We skip any pretraining with the initial velocity model, instead using it as prior information in a regularization term within the loss function. To ensure stability in the early training stages, we set a higher weight factor, $\alpha$, for this regularization term, which is gradually reduced as training progresses. This approach supports initially robust learning of the velocity model, while in later stages, the influence of the initial velocity constraint is reduced and gradually removed to allow for high resolution inverted models. However, for real data applications, we recommend retaining this regularization term even in the later training stages to counteract the potential influence of noise on the data-fitting term.

\section{Conclusions}

We proposed a new effective and practical physics-constrained deep-learning based source-independent full-waveform inversion (DL-SIFWI) method to address critical challenges in traditional FWI, including source uncertainty, amplitude dependency, and cycle skipping. By introducing a new correlation-based SIFWI objective function, we effectively mitigate source uncertainty and amplitude dependency, enhancing the robustness and accuracy of the inversion process for FWI's practical applications. The method integrates this objective function into a deep-learning framework as the loss function with a neural network (autoencoder) representation of the velocity model, which allows for an extended model search space that can help us avoid cycle skipping. We also introduce the initial model as a regularization term to speed up the convergence in the early inversion stage, since the velocity network starts with random initialization. Nevertheless, the framework converges even without initial model and relying on the random initialization of the velocity network. Through synthetic and real data experiments, we demonstrate that DL-SIFWI outperforms conventional inversion methods, even when the initial model, low-frequency data, and source wavelet information are poor. These results highlight the potential of DL-SIFWI as a powerful tool for seismic imaging, capable of delivering high-quality velocity models under challenging conditions.

\section*{Data and code availability}

The related codes for researchers to evaluate the proposed method will be uploaded to https://github.com/songc0a/DL-SIFWI after acceptance.

\section{Acknowledgement}

We thank Jilin University for its support.  We thank Viridien for providing the field data set and Geoscience Australia for providing the well-log information.We thank Statoil ASA and the Volve license partners ExxonMobil E\& P Norway AS and Bayerngas Norge AS for the release of the Volve data.

\bibliographystyle{unsrt}  
\bibliography{refs/manuscript}

\begin{thebibliography}{10}

\bibitem{tarantola1984inversion}
Albert Tarantola.
\newblock Inversion of seismic reflection data in the acoustic approximation.
\newblock {\em Geophysics}, 49(8):1259--1266, 1984.

\bibitem{brenders2007full}
AJ~Brenders and RG~Pratt.
\newblock Full waveform tomography for lithospheric imaging: results from a
  blind test in a realistic crustal model.
\newblock {\em Geophysical Journal International}, 168(1):133--151, 2007.

\bibitem{virieux2009overview}
Jean Virieux and St{\'e}phane Operto.
\newblock An overview of full-waveform inversion in exploration geophysics.
\newblock {\em Geophysics}, 74(6):WCC1--WCC26, 2009.

\bibitem{wang2009reflection}
Yanghua Wang and Ying Rao.
\newblock Reflection seismic waveform tomography.
\newblock {\em Journal of Geophysical Research: Solid Earth}, 114(B3), 2009.

\bibitem{sirgue2010thematic}
Laurent Sirgue, Olav~I Barkved, J~Dellinger, John Etgen, U~Albertin, and Jan~H
  Kommedal.
\newblock Thematic set: Full waveform inversion: The next leap forward in
  imaging at {V}alhall.
\newblock {\em First Break}, 28(4), 2010.

\bibitem{alkhalifah2016full}
Tariq~A Alkhalifah.
\newblock {\em Full Waveform Inversion in an Anisotropic World (EET 10): Where
  are the parameters hiding?}
\newblock Earthdoc, 2016.

\bibitem{luo1991wave}
Yi~Luo and Gerard~T Schuster.
\newblock Wave-equation traveltime inversion.
\newblock {\em Geophysics}, 56(5):645--653, 1991.

\bibitem{leung2006adjoint}
Shingyu Leung and Jianliang Qian.
\newblock An adjoint state method for three-dimensional transmission traveltime
  tomography using first-arrivals.
\newblock 4(1):249--266, 2006.

\bibitem{sava2004wave}
Paul Sava and Biondo Biondi.
\newblock Wave-equation migration velocity analysis. i. theory.
\newblock {\em Geophysical Prospecting}, 52(6):593--606, 2004.

\bibitem{symes2008migration}
William~W Symes.
\newblock Migration velocity analysis and waveform inversion.
\newblock {\em Geophysical prospecting}, 56(6):765--790, 2008.

\bibitem{chi2014full}
Benxin Chi, Liangguo Dong, and Yuzhu Liu.
\newblock Full waveform inversion method using envelope objective function
  without low frequency data.
\newblock {\em Journal of Applied Geophysics}, 109:36--46, 2014.

\bibitem{wu2014seismic}
Ru-Shan Wu, Jingrui Luo, and Bangyu Wu.
\newblock Seismic envelope inversion and modulation signal model.
\newblock {\em Geophysics}, 79(3):WA13--WA24, 2014.

\bibitem{xu2012full}
Sheng Xu, D~Wang, F~Chen, Yu~Zhang, and G~Lambare.
\newblock Full waveform inversion for reflected seismic data.
\newblock In {\em 74th EAGE Conference and Exhibition incorporating EUROPEC
  2012}, pages cp--293. European Association of Geoscientists \& Engineers,
  2012.

\bibitem{wu2015simultaneous}
Zedong Wu and Tariq Alkhalifah.
\newblock Simultaneous inversion of the background velocity and the
  perturbation in full-waveform inversion.
\newblock {\em Geophysics}, 80(6):R317--R329, 2015.

\bibitem{yao2020review}
Gang Yao, Di~Wu, and Shang-Xu Wang.
\newblock A review on reflection-waveform inversion.
\newblock {\em Petroleum Science}, 17(2):334--351, 2020.

\bibitem{warner2016adaptive}
Michael Warner and Llu{\'\i}s Guasch.
\newblock Adaptive waveform inversion: Theory.
\newblock {\em Geophysics}, 81(6):R429--R445, 2016.

\bibitem{guasch2019adaptive}
Llu{\'\i}s Guasch, Michael Warner, and C{\'e}line Ravaut.
\newblock Adaptive waveform inversion: Practice.
\newblock {\em Geophysics}, 84(3):R447--R461, 2019.

\bibitem{van2013mitigating}
Tristan Van~Leeuwen and Felix~J Herrmann.
\newblock Mitigating local minima in full-waveform inversion by expanding the
  search space.
\newblock {\em Geophysical Journal International}, 195(1):661--667, 2013.

\bibitem{van2015penalty}
Tristan Van~Leeuwen and Felix~J Herrmann.
\newblock A penalty method for {PDE}-constrained optimization in inverse
  problems.
\newblock {\em Inverse Problems}, 32(1):015007, 2015.

\bibitem{song2020efficient}
Chao Song and Tariq Alkhalifah.
\newblock An efficient wavefield inversion for transversely isotropic media
  with a vertical axis of symmetry.
\newblock {\em Geophysics}, 85(3):R195--R206, 2020.

\bibitem{engquist2016optimal}
Bjorn Engquist, Brittany~D Froese, and Yunan Yang.
\newblock Optimal transport for seismic full waveform inversion.
\newblock {\em arXiv preprint arXiv:1602.01540}, 2016.

\bibitem{yang2018application}
Yunan Yang, Bj{\"o}rn Engquist, Junzhe Sun, and Brittany~F Hamfeldt.
\newblock Application of optimal transport and the quadratic wasserstein metric
  to full-waveform inversion.
\newblock {\em Geophysics}, 83(1):R43--R62, 2018.

\bibitem{choi2011source}
Yunseok Choi and Tariq Alkhalifah.
\newblock Source-independent time-domain waveform inversion using convolved
  wavefields: Application to the encoded multisource waveform inversion.
\newblock {\em Geophysics}, 76(5):R125--R134, 2011.

\bibitem{wang2018microseismic}
Hanchen Wang and Tariq Alkhalifah.
\newblock Microseismic imaging using a source function independent full
  waveform inversion method.
\newblock {\em Geophysical Journal International}, 214(1):46--57, 2018.

\bibitem{choi2012application}
Yunseok Choi and Tariq Alkhalifah.
\newblock Application of multi-source waveform inversion to marine streamer
  data using the global correlation norm.
\newblock {\em Geophysical Prospecting}, 60(4):748--758, 2012.

\bibitem{plessix2006review}
R-E Plessix.
\newblock A review of the adjoint-state method for computing the gradient of a
  functional with geophysical applications.
\newblock {\em Geophysical Journal International}, 167(2):495--503, 2006.

\bibitem{karniadakis2021physics}
George~Em Karniadakis, Ioannis~G Kevrekidis, Lu~Lu, Paris Perdikaris, Sifan
  Wang, and Liu Yang.
\newblock Physics-informed machine learning.
\newblock {\em Nature Reviews Physics}, 3(6):422--440, 2021.

\bibitem{wu2023sensing}
Xinming Wu, Jianwei Ma, Xu~Si, Zhengfa Bi, Jiarun Yang, Hui Gao, Dongzi Xie,
  Zhixiang Guo, and Jie Zhang.
\newblock Sensing prior constraints in deep neural networks for solving
  exploration geophysical problems.
\newblock {\em Proceedings of the National Academy of Sciences},
  120(23):e2219573120, 2023.

\bibitem{schuster2024review}
Gerard~T Schuster, Yuqing Chen, and Shihang Feng.
\newblock Review of physics-informed machine learning inversion of geophysical
  data.
\newblock {\em Geophysics}, 89(6):1--91, 2024.

\bibitem{baydin2018automatic}
Atilim~Gunes Baydin, Barak~A Pearlmutter, Alexey~Andreyevich Radul, and
  Jeffrey~Mark Siskind.
\newblock Automatic differentiation in machine learning: a survey.
\newblock {\em Journal of Marchine Learning Research}, 18:1--43, 2018.

\bibitem{richardson2018seismic}
Alan Richardson.
\newblock Seismic full-waveform inversion using deep learning tools and
  techniques.
\newblock {\em arXiv preprint arXiv:1801.07232}, 2018.

\bibitem{zhu2021general}
Weiqiang Zhu, Kailai Xu, Eric Darve, and Gregory~C Beroza.
\newblock A general approach to seismic inversion with automatic
  differentiation.
\newblock {\em Computers \& Geosciences}, 151:104751, 2021.

\bibitem{song2023weighted}
Chao Song, Yanghua Wang, Alan Richardson, and Cai Liu.
\newblock Weighted envelope correlation-based waveform inversion using
  automatic differentiation.
\newblock {\em IEEE Transactions on Geoscience and Remote Sensing}, 61:4505011,
  2023.

\bibitem{yang2019deep}
Fangshu Yang and Jianwei Ma.
\newblock Deep-learning inversion: A next-generation seismic velocity model
  building method.
\newblock {\em Geophysics}, 84(4):R583--R599, 2019.

\bibitem{wu2019inversionnet}
Yue Wu and Youzuo Lin.
\newblock Inversionnet: An efficient and accurate data-driven full waveform
  inversion.
\newblock {\em IEEE Transactions on Computational Imaging}, 6:419--433, 2019.

\bibitem{deng2022openfwi}
Chengyuan Deng, Shihang Feng, Hanchen Wang, Xitong Zhang, Peng Jin, Yinan Feng,
  Qili Zeng, Yinpeng Chen, and Youzuo Lin.
\newblock Openfwi: Large-scale multi-structural benchmark datasets for full
  waveform inversion.
\newblock {\em Advances in Neural Information Processing Systems},
  35:6007--6020, 2022.

\bibitem{kazei2021mapping}
Vladimir Kazei, Oleg Ovcharenko, Pavel Plotnitskii, Daniel Peter, Xiangliang
  Zhang, and Tariq Alkhalifah.
\newblock Mapping full seismic waveforms to vertical velocity profiles by deep
  learning.
\newblock {\em Geophysics}, 86(5):R711--R721, 2021.

\bibitem{zhu2022integrating}
Weiqiang Zhu, Kailai Xu, Eric Darve, Biondo Biondi, and Gregory~C Beroza.
\newblock Integrating deep neural networks with full-waveform inversion:
  Reparameterization, regularization, and uncertainty quantification.
\newblock {\em Geophysics}, 87(1):R93--R109, 2022.

\bibitem{waheed2021pinntomo}
Umair~Bin Waheed, Tariq Alkhalifah, Ehsan Haghighat, Chao Song, and Jean
  Virieux.
\newblock Pinntomo: Seismic tomography using physics-informed neural networks.
\newblock {\em arXiv preprint arXiv:2104.01588}, 2021.

\bibitem{rasht2022physics}
Majid Rasht-Behesht, Christian Huber, Khemraj Shukla, and George~Em
  Karniadakis.
\newblock Physics-informed neural networks (pinns) for wave propagation and
  full waveform inversions.
\newblock {\em Journal of Geophysical Research: Solid Earth},
  127(5):e2021JB023120, 2022.

\bibitem{song2021wavefield}
Chao Song and Tariq~A Alkhalifah.
\newblock Wavefield reconstruction inversion via physics-informed neural
  networks.
\newblock {\em IEEE Transactions on Geoscience and Remote Sensing}, 60:5908012,
  2022.

\bibitem{sun2023implicit}
Jian Sun, Kristopher Innanen, Tianze Zhang, and Daniel Trad.
\newblock Implicit seismic full waveform inversion with deep neural
  representation.
\newblock {\em Journal of Geophysical Research: Solid Earth},
  128(3):e2022JB025964, 2023.

\bibitem{zhang2023multilayer}
Tianze Zhang, Jian Sun, Daniel Trad, and Kristopher Innanen.
\newblock Multilayer perceptron and bayesian neural network-based elastic
  implicit full waveform inversion.
\newblock {\em IEEE Transactions on Geoscience and Remote Sensing}, 61:1--16,
  2023.

\bibitem{yang2023fwigan}
Fangshu Yang and Jianwei Ma.
\newblock Fwigan: Full-waveform inversion via a physics-informed generative
  adversarial network.
\newblock {\em Journal of Geophysical Research: Solid Earth},
  128(4):e2022JB025493, 2023.

\bibitem{mardan2024physics}
A~Mardan and G~Fabien-Ouellet.
\newblock Physics-informed attention-based neural network for full-waveform
  inversion.
\newblock In {\em 85th EAGE Annual Conference \& Exhibition (including the
  Workshop Programme)}, volume 2024, pages 1--5. European Association of
  Geoscientists \& Engineers, 2024.

\bibitem{zhang2020high}
Zhen-dong Zhang and Tariq Alkhalifah.
\newblock High-resolution reservoir characterization using deep learning-aided
  elastic full-waveform inversion: The north sea field data example.
\newblock {\em Geophysics}, 85(4):WA137--WA146, 2020.

\bibitem{sun2023full}
Pengpeng Sun, Fangshu Yang, Hongxian Liang, and Jianwei Ma.
\newblock Full-waveform inversion using a learned regularization.
\newblock {\em IEEE Transactions on Geoscience and Remote Sensing}, 61:5920715,
  2023.

\bibitem{yao2023regularization}
Jiashun Yao, Michael Warner, and Yanghua Wang.
\newblock Regularization of anisotropic full-waveform inversion with multiple
  parameters by adversarial neural networks.
\newblock {\em Geophysics}, 88(1):R95--R103, 2023.

\bibitem{wu2017efficient}
Zedong Wu and Tariq Alkhalifah.
\newblock Efficient scattering-angle enrichment for a nonlinear inversion of
  the background and perturbations components of a velocity model.
\newblock {\em Geophysical Journal International}, 210(3):1981--1992, 2017.

\bibitem{symes2020wavefield}
William~W Symes.
\newblock Wavefield reconstruction inversion: an example.
\newblock {\em Inverse Problems}, 36(10):105010, 2020.

\bibitem{hornik1991approximation}
Kurt Hornik.
\newblock Approximation capabilities of multilayer feedforward networks.
\newblock {\em Neural networks}, 4(2):251--257, 1991.

\bibitem{leshno1993multilayer}
Moshe Leshno, Vladimir~Ya Lin, Allan Pinkus, and Shimon Schocken.
\newblock Multilayer feedforward networks with a nonpolynomial activation
  function can approximate any function.
\newblock {\em Neural networks}, 6(6):861--867, 1993.

\bibitem{dhara2022physics}
Arnab Dhara and Mrinal~K Sen.
\newblock Physics-guided deep autoencoder to overcome the need for a starting
  model in full-waveform inversion.
\newblock {\em The Leading Edge}, 41(6):375--381, 2022.

\bibitem{richardson_alan_2022}
Alan Richardson.
\newblock Deepwave.
\newblock November 2022.

\bibitem{song2020source}
Chao Song and Tariq Alkhalifah.
\newblock Source-independent efficient wavefield inversion.
\newblock {\em Geophysical Journal International}, 222(1):697--714, 2020.

\bibitem{soubaras2010variable}
Robert Soubaras and Robert Dowle.
\newblock Variable-depth streamer--a broadband marine solution.
\newblock {\em first break}, 28(12), 2010.

\bibitem{szydlik20103d}
Teresa Szydlik, Patrick Smith, Simon Way, Lars Aamodt, and Christina Friedrich.
\newblock {3D PP/PS} prestack depth migration on the {Volve} field.
\newblock 2010.

\end{thebibliography}

\end{document}